\documentclass[iop]{emulateapj}
\usepackage{natbib}
\newcommand{\myemail}{p.cargile@vanderbilt.edu}
\slugcomment{Accepted for publication in the Astrophysical Journal}
\shorttitle{Rotation Periods and Gyrochronology of Blanco~1}
\shortauthors{Cargile, James, et al.}
\begin{document}
\title{Evaluating Gyrochronology on the Zero-Age-Main-Sequence: \\Rotation Periods in the Southern Open Cluster Blanco~1 from the KELT-South Survey}
\author{P.~A. Cargile\altaffilmark{1}, D.~J. James\altaffilmark{2}, J. Pepper\altaffilmark{3,1}, R.~B. Kuhn\altaffilmark{4,5}, R. Siverd\altaffilmark{1}, K.~G. Stassun\altaffilmark{1,6}}
\altaffiltext{1}{Department of Physics and Astronomy, Vanderbilt University, Nashville, TN 37235, USA, \myemail}
\altaffiltext{2}{Cerro Tololo Inter-American Observatory, La Serena, Chile}
\altaffiltext{3}{Department of Physics, Lehigh University, Bethlehem, USA}
\altaffiltext{4}{South African Astronomical Observatory, Cape Town, South Africa}
\altaffiltext{5}{Astrophysics, Cosmology and Gravity Centre, Department of Astronomy, University of Cape Town, South Africa}
\altaffiltext{6}{Department of Physics, Fisk University, Nashville, USA}
\begin{abstract}
We report periods for 33 members of Blanco~1 as measured from KELT-South light curves, the first reported rotation periods for this benchmark zero-age-main-sequence open cluster. The distribution of these stars spans from late-A or early-F dwarfs to mid-K with periods ranging from less than a day to $\sim$8 days. The rotation period distribution has a morphology similar to the coeval Pleiades cluster, suggesting the universal nature of stellar rotation distributions. Employing two different gyrochronology methods, we find an age of 146$^{+13}_{-14}$ Myr for the cluster. Using the same techniques, we infer an age of 134$^{+9}_{-10}$ Myr for the Pleiades measured from existing literature rotation periods. These rotation-derived ages agree with independently determined cluster ages based on the lithium depletion boundary technique. Additionally, we evaluate different gyrochronology models, and quantify levels of agreement between the models and the Blanco~1/Pleiades rotation period distributions, including incorporating the rotation distributions of clusters at ages up to 1.1 Gyr. We find the Skumanich-like spin-down rate sufficiently describes the rotation evolution of stars hotter than the Sun; however, we find cooler stars rotating faster than predicted by a Skumanich-law, suggesting a mass dependence in the efficiency of stellar angular momentum loss rate. Finally, we compare the Blanco~1 and Pleiades rotation period distributions to available non-linear angular momentum evolution models. We find they require a significant mass dependence on the initial rotation rate of solar-type stars to reproduce the observed range of rotation periods at a given stellar mass, and are furthermore unable to predict the observed over-density of stars along the upper-envelope of the clusters' rotation distributions.
\end{abstract}
\section{Introduction}\label{sec.intro}
Age is a fundamental stellar parameter, however, our most commonly applied stellar chronometer (using a star's position on the HR diagram) cannot accurately place constraints on the age of a randomly-located, single, field solar-type dwarf better than a few Gyr. Therefore, it is imperative to identify and calibrate observable proxies for stellar age that are effective for main-sequence solar-type stars. One of the most important is the so-called activity-rotation-age paradigm. For a detailed description of the qualities for an effective stellar chronometer, see discussion in \citet{Barnes2007} and \citet{Soderblom2010}.

Using rotation rate as a stellar chronometer began in the 1960's, resulting in the now famous \citet{Skumanich1972} relationship which relates stellar rotation rate (specifically, $v\sin{i}$) to age, $t$, by $v\sin{i} \varpropto t^{-0.5}$. Subsequent studies have shown that although the Skumanich relationship encapsulates something fundamental about the process of stellar angular momentum loss, the actual link between a star's rotation rate and its age is a more complex, mass-dependent relationship \citep{Kawaler1988,Barnes2003a}. More recently, many groups have used extensive catalogs of stellar rotation periods for stars in open clusters of known ages to establish empirical calibrations of this age-rotation relationship \citep{Barnes2003a,Barnes2007,Mamajek2008, Meibom2009,James2010}. In addition, there are many studies that attempt to establish a theoretical framework that explains the underlying physics governing how fast a star loses its angular momentum content \citep{Barnes2010a,Barnes2010b,Matt2012,Reiners2012}. Recently, \citet{Epstein2012} found that these stellar rotation models in fact can differ in their predicted ages by as much as $\sim$30$\%$. Since uncertainties in the ages of stars (or stellar populations) used to calibrate these relationships enter linearly into their predicted ages \citep{Epstein2012}, there is a critical need for rotation period datasets from stars with known ages accurate enough to compare and contrast these models (i.e., with age uncertainties much less than $\sim$30 Myr).

Blanco~1 is a relatively young ($\sim$130 Myr; \citealt{Cargile2010b}), nearby Southern open cluster (d$\sim$200--250 pc; \citealt{vanLeeuwen2009,Platais2011}) of particular astrophysical interest due to its high Galactic latitude ($b = -79\deg$), and its comparable age to the benchmark zero-age main-sequence Pleiades open cluster. Its Galactic location also makes it an highly attractive target due its low level of field star contamination, which makes membership selection relatively straightforward \citep[e.g.,][]{Mermilliod2008a}. However, the extent of Blanco~1 on the sky -- the cluster is spread over $>$9 deg$^{2}$ \citep{Cargile2009,Platais2011} -- has made comprehensive surveys of the cluster very observationally and computationally expensive, particularly for studies requiring multiple observations (e.g., measuring rotation periods for cluster members from time-series photometric data). Nevertheless, the membership for Blanco~1 has been determined for late B-type stars \citep{Gonzalez2009} down to substellar objects \citep{Moraux2007,Mermilliod2008a,Platais2011,Casewell2012}, resulting in more than 300 currently confirmed members and an estimated $\sim$700 total members \citep{Moraux2007}. 

A key feature of Blanco~1, setting it apart as a benchmark open cluster, is that it currently is one of only eight clusters that have their ages determined using the lithium depletion boundary (LDB) technique. Although this technique is sensitive to error in distance, observational and theoretical uncertainties errors are predicted to result in age uncertainties of only 10\% \citep{Burke2004}, much less than the uncertainty compared to other more commonly applied chronometers (e.g., isochrone analysis). \citet{Cargile2010b} identified the lithium depletion boundary in Blanco~1 and determined an age of 132$\pm$24 Myr.

In this paper, we use newly-acquired Blanco 1 stellar rotation periods from the KELT-South survey to assess the accuracy and consistency of current angular momentum evolution models and gyrochronology relationships. We give an overview of the KELT-South survey (\S \ref{sec.kelts}), discuss the identification of Blanco~1 cluster members in the KELT-South database (\S \ref{sec.targets}), and provide details on our technique to determine and measure periodic variability in Blanco~1 stars (\S \ref{sec.B1rot}). We then characterize the periodic variability and rotation period distribution of solar-type stars, and report periods for 33 cluster members. These constitute the first reported rotation periods for stars this benchmark cluster. We then use this data to perform an extensive test of stellar spin-down models as well as provide a gyrochronology age for Blanco~1.
\section{Data and Analysis}
\subsection{KELT-South Survey of Blanco~1}\label{sec.kelts}
The Kilodegree Extremely Little Telescope (KELT) project is a survey designed to detect planetary transits of bright stars. KELT uses two telescopes: KELT-North in Arizona and KELT-South in Sutherland, South Africa. For full details of the survey, see \citet{Pepper2012}. KELT-South is a small-aperture (42 mm), wide-field automated telescope located at the Sutherland site of the South African Astronomical Observatory, South Africa. The telescope surveys a set of $26\deg \times 26\deg$ fields around the southern sky, obtaining $\leq$ 1$\%$ relative photometry for stars in the range of 8 $<$ V $<$ 11 mag, in order to search for transiting hot Jupiters. The survey overall provides relative photometry better than $\sim$10$\%$ for stars brighter than V$\sim$15 mag.

As part of the commissioning campaign of the KELT-South telescope, a field approximately centered on Blanco~1 ($\alpha$, $\delta$ [J2000]: $0^{h}$, $-30^{\deg}$) was observed over 90 nights from September to December 2009. Precise relative photometry was measured for $\sim$49,000 objects using a custom difference-imaging-analysis pipeline \citep[][; Siverd et al. 2013, in preparation]{Siverd2012,Pepper2012}. After outlier data rejection and red noise trend filtering \citep{Kovacs2005}, the resulting light curves for the commissioning field comprise 2,123 good-quality images from 43 separate nights, including $\sim$4500 light curves with better than 1$\%$ rms.
\subsection{Blanco~1 Membership Identification}\label{sec.targets}
\begin{figure*}
 \centering
 \includegraphics[scale=0.65,angle=0]{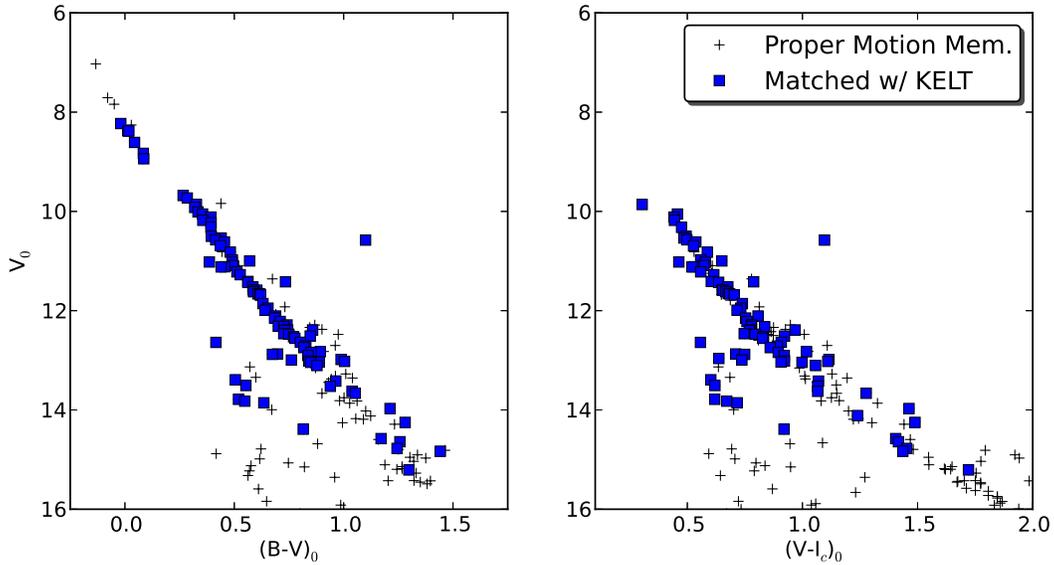} 
 \caption{
   \label{fig.MatCMD}  
   V/B$-$V and V/V$-$I$_{c}$ color-magnitude diagrams for Blanco~1 proper motion members (crosses) with matched KELT-South sources identified (squares). Reddening values used to place photometry onto intrinsic color and magnitude scales are E(B$-$V)$=$0.016 and E(V$-$I$_{c}$)$=$0.02 \citep{Cargile2009}.
   }
\end{figure*}
\begin{figure*}
 \centering
 \includegraphics[scale=0.65,angle=0]{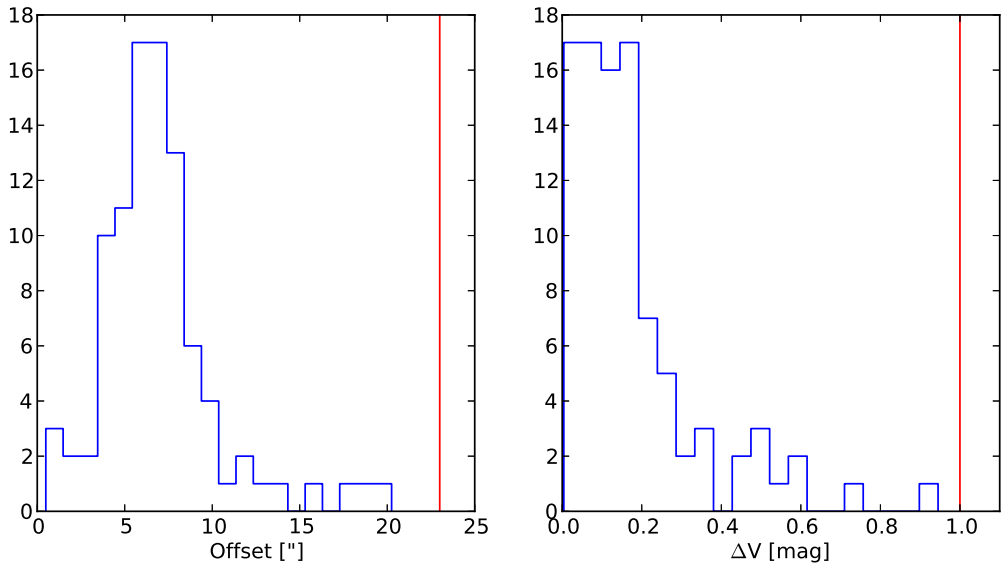} 
 \caption{
   \label{fig.MatOffset}  
   Offset distributions for the position (left) and V magnitude (right) of objects with KELT-South light curves and proper motion members of Blanco~1. The red vertical line indicates the upper limit cutoff for matching the two catalogs.
   }
\end{figure*}

In order to identify Blanco~1 members within the KELT-South commissioning data, we cross-matched the positions of the objects detected with KELT-South with a catalog of Blanco~1 high fidelity proper motion cluster members ($P_{\mu} \geq$ 5 $\%$ and $\sigma_{\mu}$ $\leq$ 2.5 mas yr$^{-1}$) identified by \citet{Platais2011}. We find 98 matches between catalogs with a search radius of 23$\arcsec$ (i.e., one KELT-South pixel), and further limit our catalog of matched Blanco~1 members to only those 94 stars that have V magnitudes estimated from the KELT instrumental magnitudes \citep[for relationship, see][]{Pepper2012} within 1 mag of the V given for the Blanco~1 cluster member in the photometric catalog of James et al. (in preparation). We plot in Fig.~\ref{fig.MatCMD} the KELT-South optical counterpart matches on the BVI$_{c}$ color-magnitude diagrams (CMD). Overall, the offset distributions (positional and measured V magnitude) of matched sources show the majority of proper motion counterparts to KELT-South sources fall within $\sim$10$\arcsec$ and have $\Delta$V$\lesssim$0.3 mag (Fig.~\ref{fig.MatOffset}). The median offset between KELT-South and Tycho positions is 7.9$\arcsec$, and because KELT-South uses a non-standard filter which is roughly equivalent to a broad V$+$I$_{c}+$R$_{c}$ band, there can be significant color terms in the calculation of V from KELT instrumental magnitudes, resulting in a 0.22 magnitude rms scatter in the KELT instrumental-to-V magnitude relationship \citep{Pepper2012}. Therefore, our matches are consistent with the expected astrometric and instrumental-to-V magnitude calibration precision of KELT-South. We note that we use the B, V, and I$_{c}$ photometry from the James et al. optical survey catalog for these matched Blanco~1 sources in all of the following analysis.

\begin{figure}
 \centering
 \includegraphics[scale=0.5,angle=0]{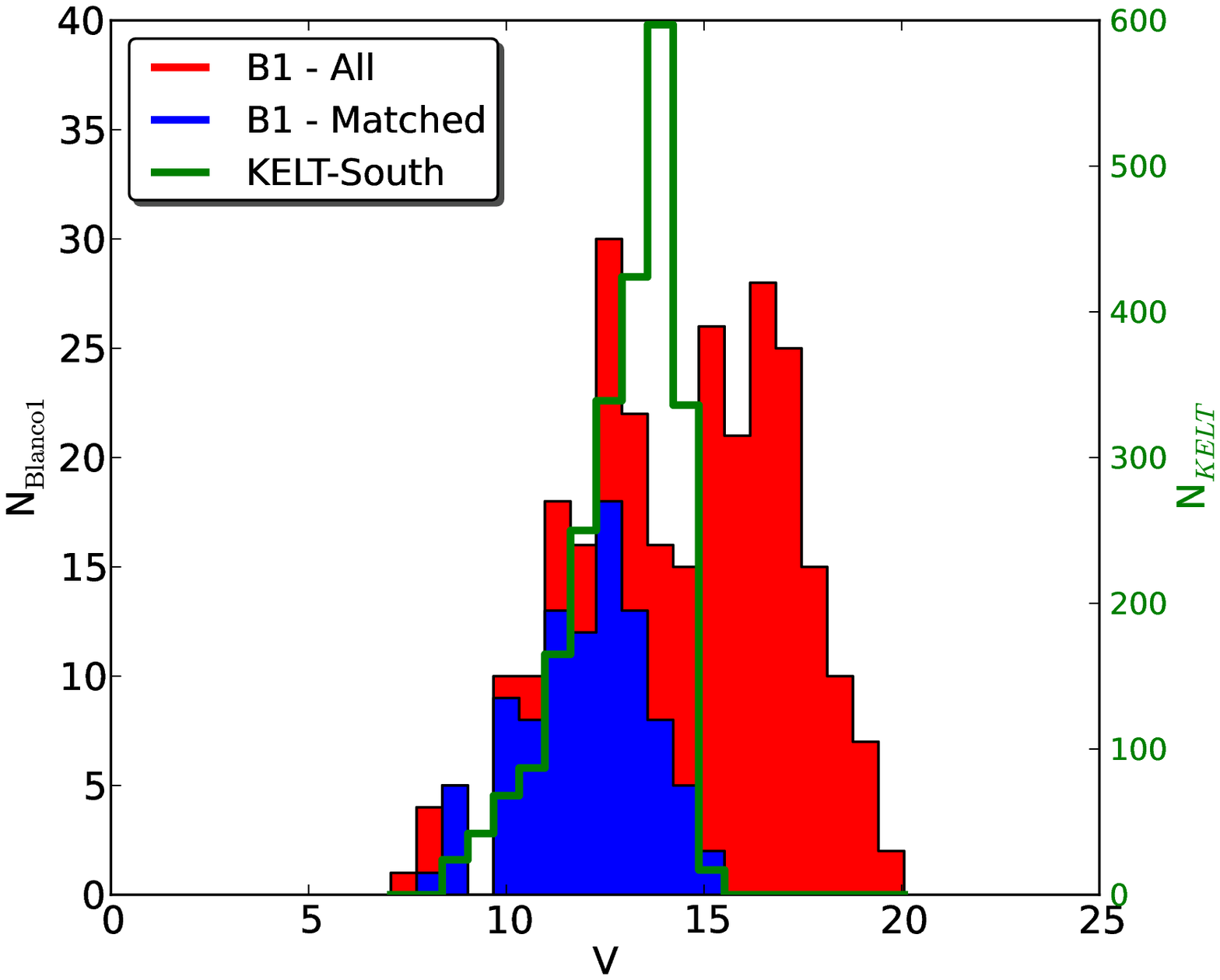} 
 \caption{
   \label{fig.MatHist}
   Brightness distribution of objects in the full KELT-South commissioning database (green), the stars identified as proper motion members of Blanco~1 from \citealt{Platais2011} (red), and the matched sources between these catalogs (blue). The vertical scale for the full proper motion and matched catalogs is provided on the left axis, while the scale for the full KELT dataset is given on the right axis.
   }
\end{figure}

 It is apparent from Fig.~\ref{fig.MatCMD} the majority of sources lie near the expected main-sequence of the cluster; however, there is clearly an increase in photometric non-members fainter than V$\sim$13 in the matched (as well as non-matched) sources. \citet{Platais2011} also points out a considerable increase in the number of contaminating field stars in the Blanco~1 proper motion membership catalog with V$>$13. We show in Fig.~\ref{fig.MatHist} a comparison of the magnitude distribution of sources in the KELT-South database and the matched/non-matched Blanco~1 proper motion members. At the bright end of our dataset, the Blanco~1 proper motion cluster members matched with KELT-South sources have a similar overall distribution to the \citeauthor{Platais2011} proper motion catalog. However, the faintness limit of the KELT-South survey significantly limits our ability to match the faintest sources in the proper motion catalog, this is seen by the abrupt decline in both the KELT-South and our matched distributions at V$\sim$14--15.

\subsection{Time Series Analysis of KELT-South Light Curves}\label{sec.B1rot}
\tabletypesize{\tiny}
\begin{deluxetable*}{p{0pt} @{ }p{0pt} @{ }p{0pt} @{}p{25pt} c c c c c c c c @{\hspace{0.25cm}}p{0.05cm} @{\hspace*{10pt}}p{5pt} @{\hspace*{-20pt}}p{5pt}}
\tablecolumns{15}
\tablewidth{7in}
\tablecaption{Blanco~1 Proper Motion Members with Periodic Signal in its KELT-South Light Curve \label{tab.KELTsources}}
\tablehead{
	\colhead{JCO\tablenotemark{a}} & \colhead{Alt.\tablenotemark{b}} & \colhead{KELT} &
	\colhead{R.A.} & \colhead{Dec.} & 
	\colhead{V} & \colhead{U$-$B} & \colhead{B$-$V} & \colhead{V$-$I$_{\rm c}$} &
	\colhead{J} & \colhead{H} & \colhead{K$_{\rm s}$} & 
	\colhead{\hspace{-7pt}Pr\tablenotemark{c}} &
	\colhead{\hspace{-7pt}Mem\tablenotemark{d}} & \colhead{\hspace{-7pt}Mem\tablenotemark{e}}\\
	\colhead{\#} & \colhead{Name} & \colhead{\#} &
	\colhead{[deg]} & \colhead{[deg]} &
	\colhead{} & \colhead{} & \colhead{} & \colhead{} &
	\colhead{} & \colhead{} & \colhead{} &
	\colhead{\hspace{-7pt}$\mu$} & 
	\colhead{\hspace{-7pt}RV} & \colhead{\hspace{-7pt}Photo}\\
}
\startdata
\nodata\hspace*{-1.5cm} & ZS226   &  02619  &   2.0079583  &  -30.0316917 &  9.780$\pm$0.001  &  0.020$\pm$0.020 &  0.301$\pm$0.001 &       \nodata      &  9.19$\pm$0.03 &  9.07$\pm$0.02 &  9.04$\pm$0.02 & 98 &  M & M \\  
1193\hspace*{-1.5cm} & ZS166   &  02901  &   1.4290000  &  -29.9606056 &  9.912$\pm$0.001  &  0.010$\pm$0.020 &  0.343$\pm$0.001 &   0.324$\pm$0.001  &  9.31$\pm$0.02 &  9.17$\pm$0.03 &  9.12$\pm$0.02 & 96 &  M & M \\  
0635\hspace*{-1.5cm} & ZS90    &  03534  &   0.8516250  &  -29.8137083 & 10.665$\pm$0.001  & -0.055$\pm$0.002 &  0.471$\pm$0.001 &   0.557$\pm$0.002  &  9.75$\pm$0.03 &  9.50$\pm$0.03 &  9.46$\pm$0.02 & 97 &  M & M \\  
0995\hspace*{-1.5cm} & ZS134   &  07000  &   1.2050000  &  -30.0146806 & 11.141$\pm$0.001  & -0.070$\pm$0.004 &  0.508$\pm$0.002 &   0.592$\pm$0.002  & 10.14$\pm$0.02 &  9.90$\pm$0.02 &  9.82$\pm$0.02 & 52 &  M & M \\  
0561\hspace*{-1.5cm} & ZS84    &  09120  &   0.7950417  &  -30.1802472 & 11.325$\pm$0.002  &  0.007$\pm$0.004 &  0.543$\pm$0.003 &   0.635$\pm$0.003  & 10.28$\pm$0.02 & 10.04$\pm$0.03 &  9.97$\pm$0.02 & 87 &  M & M \\  
1038\hspace*{-1.5cm} & ZS138   &  09841  &   1.2451667  &  -30.1615583 & 11.461$\pm$0.001  &  0.023$\pm$0.004 &  0.578$\pm$0.002 &   0.624$\pm$0.002  & 10.42$\pm$0.02 & 10.13$\pm$0.02 & 10.08$\pm$0.02 & 95 &  M & M \\  
1240\hspace*{-1.5cm} & W108    &  10120  &   1.4850000  &  -28.9363361 & 11.478$\pm$0.002  &  0.039$\pm$0.004 &  0.578$\pm$0.003 &   0.656$\pm$0.003  & 10.42$\pm$0.02 & 10.15$\pm$0.02 & 10.12$\pm$0.02 & 84 &  M & M \\  
1037\hspace*{-1.5cm} & W89     &  10518  &   1.2443333  &  -29.5632472 & 11.571$\pm$0.002  &  0.033$\pm$0.005 &  0.601$\pm$0.003 &   0.696$\pm$0.003  & 10.46$\pm$0.02 & 10.16$\pm$0.03 & 10.09$\pm$0.02 & 96 &  M & M \\  
0538\hspace*{-1.5cm} & W52     &  11269  &   0.7771667  &  -29.3622222 & 11.636$\pm$0.002  &  0.055$\pm$0.005 &  0.619$\pm$0.003 &   0.675$\pm$0.003  & 10.54$\pm$0.02 & 10.26$\pm$0.02 & 10.21$\pm$0.02 & 89 &  M & M \\  
1248\hspace*{-1.5cm} & ZS176   &  09392  &   1.4960417  &  -29.6512917 & 11.669$\pm$0.002  &  0.031$\pm$0.006 &  0.604$\pm$0.004 &   0.688$\pm$0.003  & 10.51$\pm$0.02 & 10.29$\pm$0.02 & 10.19$\pm$0.02 & 97 &  M & M \\  
1315\hspace*{-1.5cm} & ZS182   &  11129  &   1.5681250  &  -30.0991833 & 11.705$\pm$0.002  &  0.086$\pm$0.007 &  0.624$\pm$0.004 &   0.698$\pm$0.004  & 10.61$\pm$0.03 & 10.31$\pm$0.03 & 10.25$\pm$0.02 & 97 &  M & M \\  
1591\hspace*{-1.5cm} & ZS201   &  12182  &   1.7591250  &  -30.3549722 & 11.712$\pm$0.001  &  0.031$\pm$0.006 &  0.633$\pm$0.003 &   0.704$\pm$0.003  & 10.56$\pm$0.02 & 10.29$\pm$0.02 & 10.22$\pm$0.02 & 92 &  M & M \\  
2654\hspace*{-1.5cm} & M330    &  09673  & 359.2209167  &  -29.8608444 & 11.734$\pm$0.003  &  0.051$\pm$0.006 &  0.636$\pm$0.004 &   0.724$\pm$0.005  & 10.59$\pm$0.03 & 10.28$\pm$0.02 & 10.24$\pm$0.02 & 97 &  M & M \\  
1322\hspace*{-1.5cm} & M340    &  13164  &   1.5740417  &  -29.3112361 & 12.040$\pm$0.002  &  0.137$\pm$0.009 &  0.657$\pm$0.004 &   0.736$\pm$0.004  & 10.84$\pm$0.02 & 10.52$\pm$0.02 & 10.44$\pm$0.02 & 64 &  M & M \\  
2763\hspace*{-1.5cm} & M346    &  10374  & 359.4677500  &  -30.0883417 & 12.157$\pm$0.003  &  0.106$\pm$0.009 &  0.705$\pm$0.005 &   0.827$\pm$0.004  & 10.84$\pm$0.02 & 10.48$\pm$0.02 & 10.40$\pm$0.02 & 95 &  M & M \\  
0462\hspace*{-1.5cm} & M337    &  18230  &   0.7032917  &  -29.3117250 & 12.337$\pm$0.003  &  0.247$\pm$0.013 &  0.757$\pm$0.005 &   0.799$\pm$0.005  & 11.03$\pm$0.02 & 10.67$\pm$0.02 & 10.59$\pm$0.02 & 91 &  M & M \\  
1862\hspace*{-1.5cm} & ZS218   &  16802  &   1.9570833  &  -30.4897778 & 12.364$\pm$0.003  &  0.228$\pm$0.013 &  0.717$\pm$0.005 &   0.798$\pm$0.005  & 11.05$\pm$0.03 & 10.68$\pm$0.02 & 10.62$\pm$0.02 & 89 &  M & M \\  
0223\hspace*{-1.5cm} & ZS58    &  16335  &   0.4435833  &  -29.7774222 & 12.374$\pm$0.003  &  0.198$\pm$0.013 &  0.743$\pm$0.006 &   0.856$\pm$0.005  & 10.86$\pm$0.02 & 10.49$\pm$0.03 & 10.42$\pm$0.02 & 96 &  M & M \\  
0493\hspace*{-1.5cm} & ZS76    &  13762  &   0.7349167  &  -30.0791028 & 12.438$\pm$0.003  &  0.417$\pm$0.017 &  0.874$\pm$0.007 &   0.988$\pm$0.004  & 10.76$\pm$0.02 & 10.28$\pm$0.02 & 10.16$\pm$0.02 & 96 &  SB2 & M \\  
0687\hspace*{-1.5cm} & ZS102   &  19594  &   0.8903750  &  -30.2622417 & 12.524$\pm$0.003  &  0.222$\pm$0.014 &  0.765$\pm$0.006 &   0.815$\pm$0.005  & 11.16$\pm$0.03 & 10.79$\pm$0.02 & 10.73$\pm$0.02 & 95 &  M & M \\  
3025\hspace*{-1.5cm} & M347    &  20765  & 359.8263333  &  -30.1712417 & 12.606$\pm$0.003  &  0.343$\pm$0.016 &  0.793$\pm$0.006 &   0.847$\pm$0.005  & 11.20$\pm$0.02 & 10.85$\pm$0.02 & 10.77$\pm$0.02 & 96 &  M & M \\  
0149\hspace*{-1.5cm} & ZS262   &  24789  &   0.3450833  &  -29.8434972 & 12.686$\pm$0.001  &  0.275$\pm$0.024 &  0.818$\pm$0.008 &   0.928$\pm$0.002  & 11.21$\pm$0.02 & 10.75$\pm$0.02 & 10.64$\pm$0.02 & 87 &  M & M \\  
0749\hspace*{-1.5cm} & ZS106   &  20006  &   0.9727083  &  -29.6426111 & 12.689$\pm$0.004  & -0.145$\pm$0.010 &  0.432$\pm$0.006 &   0.577$\pm$0.007  & 11.76$\pm$0.02 & 11.54$\pm$0.03 & 11.51$\pm$0.03 & 32 &  \nodata & NM\\  
0275\hspace*{-1.5cm} & BLX7    &  20495  &   0.5033333  &  -29.9881611 & 12.762$\pm$0.004  &  0.384$\pm$0.022 &  0.843$\pm$0.008 &   0.911$\pm$0.006  & 11.21$\pm$0.02 & 10.79$\pm$0.02 & 10.69$\pm$0.02 & 87 &  M & M \\  
2774\hspace*{-1.5cm} & M349    &  23722  & 359.4875417  &  -29.3611000 & 12.877$\pm$0.004  &  0.427$\pm$0.023 &  0.909$\pm$0.008 &   1.038$\pm$0.006  & 11.09$\pm$0.02 & 10.64$\pm$0.02 & 10.50$\pm$0.02 & 82 &  M & M \\  
2119\hspace*{-1.5cm} & PM1636  &  26823  &   2.3495417  &  -30.6446528 & 12.922$\pm$0.004  &     \nodata      &  0.714$\pm$0.006 &   0.730$\pm$0.006  & 11.67$\pm$0.02 & 11.38$\pm$0.03 & 11.32$\pm$0.02 & 35 &  \nodata & NM\\  
0333\hspace*{-1.5cm} & ZS45    &  23443  &   0.5772917  &  -29.8523528 & 12.955$\pm$0.006  &  0.591$\pm$0.040 &  0.853$\pm$0.013 &   0.941$\pm$0.008  & 11.38$\pm$0.02 & 10.97$\pm$0.02 & 10.84$\pm$0.02 & 93 &  M & M \\  
0374\hspace*{-1.5cm} & ZS54    &  22952  &   0.6174583  &  -30.0787528 & 13.033$\pm$0.005  &  0.787$\pm$0.046 &  1.004$\pm$0.011 &   1.135$\pm$0.007  & 11.14$\pm$0.02 & 10.69$\pm$0.03 & 10.51$\pm$0.02 & 95 &  M & M \\  
0786\hspace*{-1.5cm} & ZS112   &  24396  &   1.0167500  &  -29.9740083 & 13.063$\pm$0.005  &  0.454$\pm$0.030 &  0.857$\pm$0.010 &   0.942$\pm$0.008  & 11.46$\pm$0.02 & 11.06$\pm$0.03 & 10.92$\pm$0.02 & 94 &  M & M \\  
2075\hspace*{-1.5cm} & ZS243   &  27884  &   2.2833333  &  -29.8078083 & 13.084$\pm$0.004  &     \nodata      &  0.904$\pm$0.008 &   0.927$\pm$0.006  & 11.45$\pm$0.02 & 11.04$\pm$0.02 & 10.92$\pm$0.02 & 93 &  SB1 & M \\  
1329\hspace*{-1.5cm} & ZS181   &  28289  &   1.5794583  &  -30.3729139 & 13.091$\pm$0.003  &  0.395$\pm$0.022 &  0.865$\pm$0.006 &   1.018$\pm$0.004  & 11.35$\pm$0.02 & 10.88$\pm$0.02 & 10.79$\pm$0.02 & 92 &  SB1 & M \\  
1553\hspace*{-1.5cm} & PM5376  & 28016   &   1.7315417  &  -29.0314472 & 13.155$\pm$0.005  &  0.514$\pm$0.031 &  0.894$\pm$0.010 &   1.077$\pm$0.007  & 11.36$\pm$0.03 & 10.94$\pm$0.03 & 10.77$\pm$0.03 & 15 &  \nodata & M \\  
1159\hspace*{-1.5cm} & ZS154   &  43425  &   1.3815833  &  -30.3476556 & 13.469$\pm$0.007  &  0.654$\pm$0.057 &  0.980$\pm$0.016 &   1.090$\pm$0.010  & 11.66$\pm$0.02 & 11.14$\pm$0.03 & 11.04$\pm$0.02 & 90 &  M & M \\  
2698\hspace*{-1.5cm} & PM3007  & 35511   & 359.3627083  &  -30.0900417 & 13.714$\pm$0.008  &  0.603$\pm$0.065 &  1.069$\pm$0.019 &   1.297$\pm$0.011  & 11.50$\pm$0.02 & 10.98$\pm$0.02 & 10.80$\pm$0.02 & 88 &  \nodata & M \\  
0539\hspace*{-1.5cm} & ZS88    &  36902  &   0.7776250  &  -29.7198556 & 14.021$\pm$0.011  &     \nodata      &  1.228$\pm$0.033 &   1.483$\pm$0.014  & 11.69$\pm$0.03 & 11.07$\pm$0.02 & 10.92$\pm$0.02 & 82 &  M & M \\  
\enddata
\tablecomments{Object positions are epoch J2000.}
\tablenotetext{a}{JCO identifier from the James et al. optical catalog.}
\tablenotetext{b}{Alternative names from extant catalogs: ZS = \citet{deEpstein1985}, W = \citet{Westerlund1988}, M = \citet{Mermilliod2008a}, BLX = \citet{Micela1999a}, and PM = \citet{Platais2011}.}
\tablenotetext{c}{Proper motion probablity from \citet{Platais2011}.}
\tablenotetext{d}{Radial velocity membership: M = radial velocity single cluster member, SB2 = double lined spectroscopic binary, SB1 = RV variable.}
\tablenotetext{e}{Photometric membership based on optical color-magnitude diagrams: M = photometric member, NM = photometric non-member.}
\end{deluxetable*}
For the 94 Blanco~1 proper motion members that we have matched with KELT-South sources, we analyze their KELT-South light curves to identify and measure periodic variability. We use a Lomb-Scargle Periodogram (LSP) \citep{Press1989}\footnote{Python code available at http://www.astropython.org} to identify periodic variability in each KELT light curve. The significance of identified peaks in each LSP is tested by using a Monte Carlo simulation of random permutations of each light curve. We simulated 10000 random permutations of each KELT light curve, randomizing the photometric values while keeping the timestamps fixed. Applying the LSP to each simulated random light curve, we determine the height distribution for the tallest peaks in the LSP for these random light curves. We then compare the height of the Blanco~1 star's LSP to this distribution to determine its false-alarm probability (FAP). We consider stars with significant periodic variability to have FAP $\leq$ 0.01, i.e., that each star's LSP has a peak with a height that is only seen per chance in less than 1\% of the 10000 randomized light curves. Of the 94 total Blanco~1 KELT-South matched sources, 40 stars had significant peaks in their LSP using this metric. 

We visually inspected each remaining LSP to reject any with known systematics in the KELT-South dataset. This includes significant periodic variability at known aliases (e.g., at day-integer periods), as well as identifying any nearby stars in the KELT-South images observed with similar variability, indicating blending contamination. After this manual filtering, we are left with 35 stars with significant and reliably measured periodic variability (Table \ref{tab.KELTsources}). Period errors for these stars were determined by calculating a 1$\sigma$ confidence level on the measured period using the {\em post mortem} Schwarzenberg-Czerny method \citep{Schwarzenberg-Czerny1991}. 
\section{Results}\label{sec.Results}
\subsection{Periodic Variability in Blanco~1}\label{sec.B1PV}
The majority of Blanco~1 stars we identify as having periodic variability in their light curves are solar- and late-type stars (late-F to late-K spectral type), apart from two variable high-mass stars (see \S \ref{subsec.HMV}). The variability seen in young, rapidly rotating stars has long been thought to be due to photospheric heterogeneities, such as magnetically induced starspots \citep[for early discussions on spot induced brightness variations on active stars, see][]{Pickering1881,Kron1947,Krzeminski1969,Evans1971}. Therefore, by accurately measuring the period of their variability, we are able to sample the rotation-rate distribution of stars in Blanco~1. Overall, our sample contains rotation periods from $\sim$0.1--7 days measured from stars with B$-$V$\sim$ 0.4--1.25 mag.
\subsection{Improved Membership for Blanco~1 Using Periodic Variability}\label{sec.membership}
Open cluster membership is typically established by constructing CMDs and identifying objects lying near the expected cluster (pre-/post-) main-sequence. Stars identified as belonging to a stellar population using this method have a probability of membership dependent on the contamination level of background and foreground objects. Therefore, additional membership criteria are commonly applied to CMDs in order to reduce the contamination level of non-member interlopers. These criteria are usually based on distinguishing cluster members by their kinematics (radial velocities, proper motions), or for young open clusters, signatures of youth not typically seen in older field stars (heightened levels of magnetic activity, photospheric Li abundance, rapid rotation). Due to the nature of each of these individual criteria, these membership properties have varying levels of field star false-positive contamination as well as biases based on their effectiveness for different colors and/or magnitude ranges \citep[e.g., see][]{Cargile2010}. 

\begin{figure*}
 \centering
 \includegraphics[scale=0.60,angle=0]{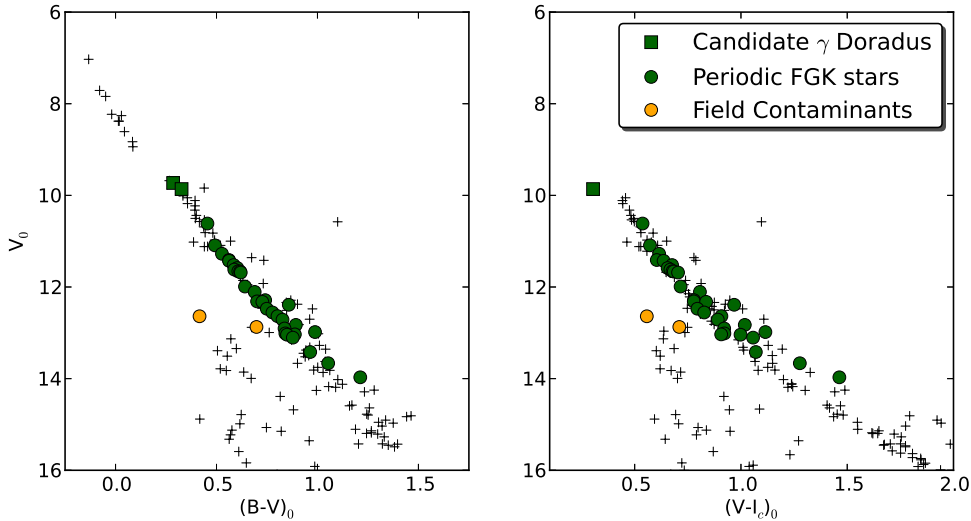} 
 \caption{
   \label{fig.PerCMD}  
   V/B$-$V and V/V$-$I$_{c}$ color-magnitude diagrams for Blanco~1. Plotted are Blanco~1 proper motion members (crosses), Blanco~1 members with periodic variability in their KELT-South light curves (green), and contaminant field stars (yellow). HD~91 and HD~343, the two potential $\gamma$ Dor pulsating stars, are identified with square symbols at the bluest part of the cluster sequence.
   }
\end{figure*}

As stated in \S \ref{sec.targets}, \citet{Platais2011} identifies non-member contamination that increases significantly for stars fainter than V$\sim$13 in their proper motion catalog. Using this dataset and an empirically determined cluster sequence based on the similarly aged Pleiades \citep{Stauffer2007}, we calculate that the \citeauthor{Platais2011} Blanco~1 proper motion catalog contains 242/281 stars that fall within their photometric uncertainties of the expected cluster sequence (taking into account that binaries can lie up to 0.75 mag above the single stars loci), resulting in an estimated contamination rate of $\sim$14\% (39/281). However, the clear majority of the 35 Blanco~1 proper motion members with detected periodic variability clearly delineate the cluster sequence in the B$-$V and V$-$I$_{c}$ CMDs (Fig.~\ref{fig.PerCMD}). We identify 33/35 objects showing periodic variability fall along the Blanco~1's sequence, with two objects lying significantly away from the cluster sequence likely being contaminating background/foreground sources and not associated with Blanco~1. This suggests that identification of periodic variability improves the identification of Blanco~1 stars down to only a 6\% (2/35) probability of contamination, and can be exploited to easily and reliably used to identify Blanco~1 cluster members. This new high-fidelity membership catalog based on stars that show periodic variability and are proper motion members provides an excellent dataset to test the accuracy of the measured properties of Blanco~1. For the remainder of our Blanco~1 variability analysis we remove the 2 photometric non-member sources, leaving a total of 33 Blanco~1 proper motion and photometric cluster members with periodic variability in their KELT light curves (Table \ref{tab.Periods}).
\begin{deluxetable}{l c c}
\tablecolumns{3}
\tablewidth{0pc}
\tablecaption{Measured Perodicity for Confirmed Blanco~1 Members \label{tab.Periods}}
\tablehead{
	\colhead{JCO}  & \colhead{Period} & \colhead{Amp.\tablenotemark{a}}\\
	\colhead{\#}   & \colhead{[d]} & \colhead{}
}
\startdata
ZS226     & 0.307$\pm$0.001 & 0.009 \\  
1193      & 1.417$\pm$0.040 & 0.012 \\  
0635      & 1.692$\pm$0.087 & 0.003 \\  
0995      & 1.347$\pm$0.008 & 0.010 \\  
0561      & 1.811$\pm$0.082 & 0.007 \\  
1038      & 0.125$\pm$0.001 & 0.014 \\  
1240      & 3.169$\pm$0.227 & 0.026 \\  
1037      & 2.872$\pm$0.048 & 0.019 \\  
0538      & 3.384$\pm$0.205 & 0.015 \\  
1248      & 0.592$\pm$0.011 & 0.012 \\  
1315      & 5.084$\pm$1.437 & 0.015 \\  
1591      & 1.695$\pm$0.058 & 0.010 \\  
2654      & 3.728$\pm$0.168 & 0.011 \\  
1322      & 4.527$\pm$1.051 & 0.016 \\  
2763      & 2.756$\pm$0.148 & 0.014 \\  
0462      & 5.528$\pm$0.357 & 0.017 \\  
1862      & 5.253$\pm$0.524 & 0.018 \\  
0223      & 2.353$\pm$0.038 & 0.017 \\  
0493      & 3.148$\pm$0.165 & 0.016 \\  
0687      & 5.496$\pm$0.657 & 0.019 \\  
3025      & 5.762$\pm$0.547 & 0.024 \\  
0149      & 0.408$\pm$0.001 & 0.042 \\  
0275      & 6.174$\pm$1.047 & 0.025 \\  
2774      & 5.869$\pm$0.933 & 0.043 \\  
0333      & 3.574$\pm$0.171 & 0.034 \\  
0374      & 5.084$\pm$0.916 & 0.021 \\  
0786      & 6.557$\pm$0.339 & 0.096 \\  
2075      & 6.743$\pm$0.601 & 0.079 \\  
1329      & 0.341$\pm$0.001 & 0.050 \\  
1553      & 6.890$\pm$0.615 & 0.025 \\  
1159      & 6.512$\pm$0.794 & 0.070 \\  
2698      & 5.593$\pm$0.499 & 0.043 \\  
0539      & 3.803$\pm$0.676 & 0.058 \\  
\enddata
\tablenotetext{a}{The peak-to-peak median amplitude of the variability in \\KELT-South instrumental magnitudes.}
\end{deluxetable}
\subsection{Potential $\gamma$ Doradus Blanco~1 Stars}\label{subsec.HMV}
\begin{figure}
 \centering
 \includegraphics[scale=0.4,angle=0]{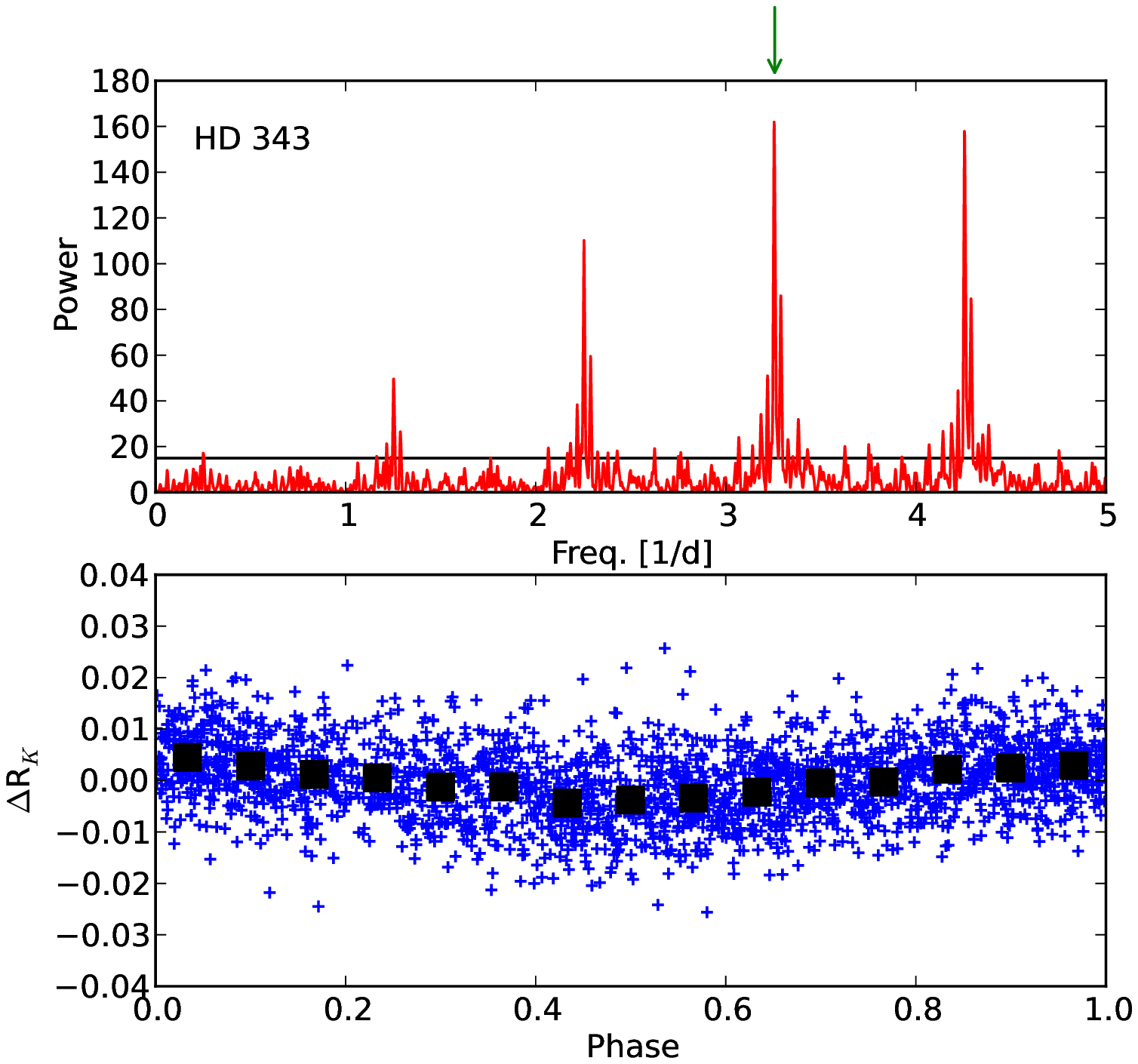}
 \includegraphics[scale=0.4,angle=0]{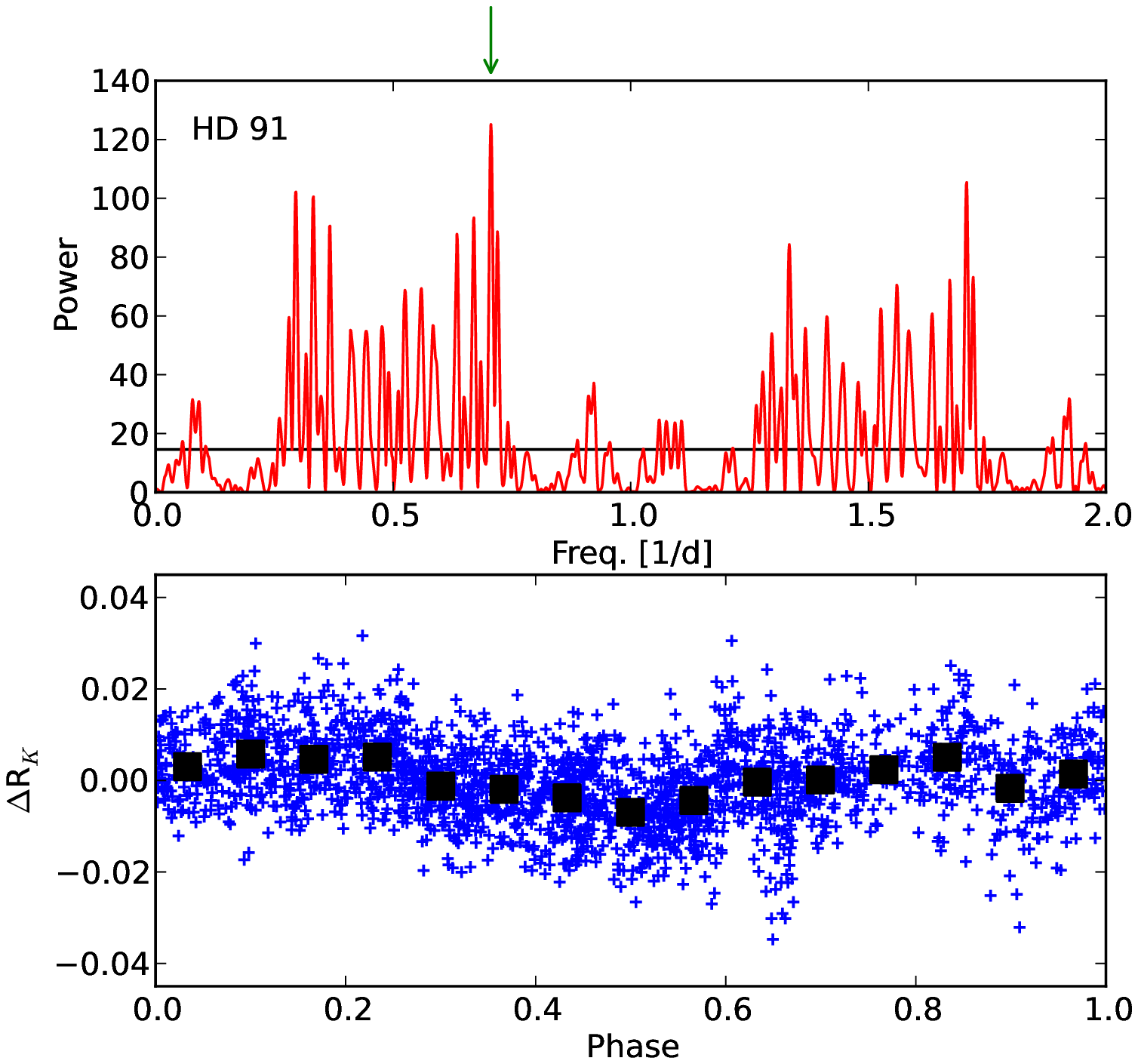}
 \caption{
 	\label{fig.HMV}  
 	Lomb-Scargle periodograms (red) and phased KELT-South light curves (blue) for two potential $\gamma$ Dor variable stars HD~343 (top panels) and HD~91 (bottom panels). Light curves have been phased to their detected periods (Period = 0.307 d for HD~343, and Period = 1.417 d for HD91) based on the tallest peak in their respective periodograms (green arrow at top of periodograms). The brightness of the stars is plotted in relative KELT R$_{\rm K}$ instrumental magnitude, and black squares show the binned light curve based on 15 equal sized phase-bins. The power at a FAP$=$0.1\% is indicated as a horizontal line in the periodograms.
   }
\end{figure}

We find that HD~343 (ZS~226) and HD~91 (ZS~166), two late-A or early-F Blanco~1 dwarfs (B$-$V$=$0.301 and 0.324, respectively), show strong signatures of periodicity in their KELT-South light curve (Fig.~\ref{fig.HMV}). We measure their variability to have clear periodicities at 0.307 and 1.417 days for HD~343 and HD~91, respectively. Some suggested mechanisms for periodic variability of late-A/early-F dwarfs are ellipsoidal distortions due to a close binary, surface features causing brightness modulations, and/or $\delta$ Sct or $\gamma$ Dor pulsations \citep[see][ and reference therein]{Henry2011}. \citet{Gonzalez2009} included these two stars as part of a long term spectroscopic survey looking for binaries in Blanco~1. These authors classified HD~91 as radial velocity constant ($\sigma_{RV}=$0.5 km s$^{-1}$ over 2 observations), and HD~343 classified as tentatively radial velocity constant ($\sigma_{RV}=$5.7 km s$^{-1}$ over 3 observations). Also, HD~91 was observed but not detected in a deep XMM-Newton survey of Blanco~1 \citep{Pillitteri2003}, and neither HD~343 nor HD~91 have excess UV emission in GALEX \citep{Martin2005}. Lower-mass, solar-type Blanco~1 members are observed to emit strongly at high energies indicative of young, rapidly rotating stars \citep{Cargile2009,Findeisen2011}, and the dearth of X-ray/UV excess in HD~91 and HD~343 suggests that they do not harbor unseen companions that could lead to ellipsoidal tidal distortions. Together, the lack of significant radial velocity variation in these stars and non-detections for excess X-ray/UV emission, gives strong evidence that they are not probable ellipsoidal variables. In addition, the late-A or early-F spectral types of these two stars, and the high level of coherence in the light curves over many periodic cycles, does not support surface inhomogeneities (e.g., starspots) causing their observed variability. Therefore, the period of HD~343 and HD~91 (0.301$\pm$0.001 d and 1.417$\pm$0.040 d, respectively), the observational evidence against the variability due to ellipsoidal tidal distortions and surface features, along with their proximity to the instability strip on the CMD, suggest strongly that these objects are undergoing pulsation, and are most likely $\gamma$ Dor pulsators as they fall on the low-mass end of the main-sequence instability strip. However, follow up spectroscopic and multi-band photometric observations of HD~343 and HD~91 are needed to confidently identify pulsations as the cause of their photometric variability.
\subsection{Comparing Stellar Rotation in Blanco~1 and coeval Pleiades}\label{sec.compPle}
The age of Blanco~1 and the Pleiades are determined to be very similar (within 10\%) based on the distribution of stellar activity \citep{Cargile2009}, upper-main-sequence turn-off age \citep[][; James et al. in prep]{Platais2011}, pre-main-squence contraction age (James et al. in prep). Further evidence comes from their distribution of stellar and sub-stellar Li abundances and subsequent modeling of their LDBs \citep{Jeffries1999a,Cargile2010b}. Apart from an apparent lack of high-mass B-dwarf members in Blanco~1, the overall mass- and luminosity-functions for these two clusters are similar \citep{Moraux2007,Platais2011}. Nevertheless, \citet{Moraux2007} calculated that Blanco~1 is less dense and half as massive as the Pleiades ($\sim$30 stars/pc$^{2}$ with 410 M$_{\odot}$ total mass versus $\sim$65 stars/pc$^{2}$ and 735 M$_{\odot}$, respectively). Blanco~1 also has a unique abundance pattern suggesting it was formed out of material not well mixed with the ISM of the Galactic disk \citep{Ford2005}, and has an atypical Galactic position for a thin-disk Galactic open cluster, i.e., $\sim$250 pc below the Galactic disk. The Pleiades, on the other hand, is a more ``typical'' Galactic open cluster, having a solar-like abundance pattern \citep{Soderblom2009} and is located within 100 parsecs of the Galactic plane \citep{Soderblom2005}. 

\begin{figure*}
 \centering
 \includegraphics[scale=0.60,angle=0]{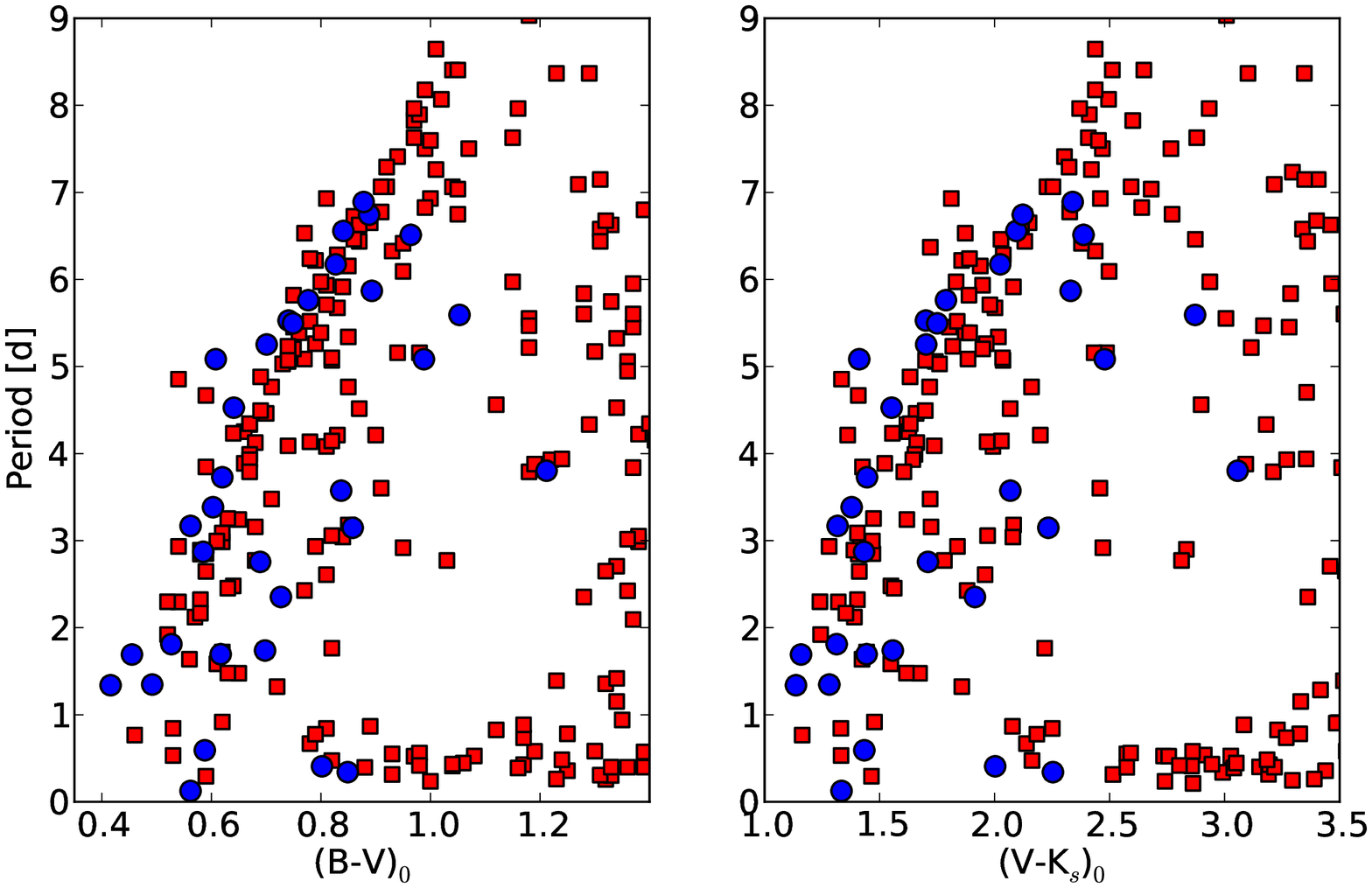} 
 \caption{
   \label{fig.ColPerBlanco1Ple}
   Color-period diagrams for KELT-South rotation periods in Blanco~1 (blue circles) and HATnet data for the Pleiades (red squares). Colors are corrected for reddening using each cluster's respective E(B$-$V) and E(V$-$K$_{s}$) values. Photometry and rotation periods for the Pleiades stars are taken from \citet{Hartman2010}. Due to the KELT faintness limit, the Blanco~1 data becomes incomplete redder than B$-$V$\sim$1.0 and V$-$K$_{s} \sim$2.75. 
   }
\end{figure*}

In Fig.~\ref{fig.ColPerBlanco1Ple}, we plot the distribution of rotation periods for Blanco~1 and the Pleiades. The Pleiades rotation periods are taken from the HATnet survey of the cluster \citep{Hartman2010}. These two coeval clusters generally appear to have similar overall rotation period distribution morphologies, namely, a well defined upper-envelopes of slower rotators (``I-sequence"), and a distribution of more rapid rotators (``gap-stars") that extend down to stars with rotation periods at or near breakup speeds (``C-sequence"). The preceding I-, C-, and gap terminology is on discussion in \citet{Barnes2003a}.

In order to determine a more quantitative measure of the similarity between these two cluster's color-period distributions, we determine the probability that the Blanco~1 rotation periods we measure are drawn from the larger Pleiades distribution. First we determine a metric of similarity between the two color-period distributions by deriving the probability density function (PDF) for each dataset individually using a kernel density estimation (KDE). Unlike histograms, where bin placement can cause significant qualitative differences, a KDE uses a kernel function to smooth out the contribution of each data point to a joint PDF, thus generating a continuous estimation of the probability of a parameter value given the overall distribution of data points \citep[for details, see][]{Silverman1986}. We determine this kernel for our case using an automatic bivariate bandwidth determination based on ``Scott's Rule" \citep{Scott1992}. 

We multiply the PDFs determined for each cluster then integrate this product over the full color and period parameter space, resulting in a metric describing the overall correlation between the Blanco~1 and Pleiades distributions. To find the significance of this metric, we use a Monte Carlo simulation of 1 million randomly drawn subsamples of 33 Pleiades data points (same number as the Blanco~1 distribution), and calculate the integrated PDF product of these subsamples and the Pleiades distribution. We repeat this Monte Carlo process by deriving this metric for 1 million randomly selected subsample of 33 data points with period and color values taken from the range observed in Pleiads. The first Monte Carlo simulation allows us to determine the probability for measuring this metric given a true subsample of the Pleiades distribution (P1), the second simulation results in a probability of deriving this metric from 33 random data points over the same parameter space (P2). The integrated probability product for Blanco~1 with the Pleiades PDF results in a correlation metric with a p-value of P1$=$0.0018, and similarly the product of Blanco~1's metric with the random distribution resulted in a p-value of P2$=$0.000035. Therefore, from these simulations we determine the Blanco~1 distribution is $\sim$50 times more likely to be drawn from the Pleiades then being drawn from a random distribution over the same parameter space.

This quantitative evidence for the similarity of the color-period distributions for Blanco~1 and the Pleiades clusters, including the range of periods at any given stellar color and the relative densities of data points, suggests that although these two coeval clusters most likely had very different formation environments and evolutionary histories, these factors do not seem to affect their overall rotation period distribution morphology. Thus, the processes that govern the variation seen in the rotation rate of individual member stars (e.g., tidally locked binary systems, circumstellar-disk evolution, etc.) must occur with similar frequency and duration in both clusters. We note this is only true if the probability of detecting rotation periods in each cluster is similar for both datasets (e.g., similar sensitivity of each dataset to the observed range in periods), however, the rotation period measurements for both clusters are well within the viable periodicity and brightness range of their respective photometric surveys (KELT-South for Blanco~1 and HATnet for the Pleiades).
\section{Testing Models of Angular Momentum Evolution and Gyrochronology with Blanco~1}\label{sec.gyroch}
\subsection{Empirical ``I-sequence'' Rotation-Age Relationships}\label{sec.Iseq}
In an effort to derive gyrochronology ages for main-sequence solar-type stars, \citet[B03;][]{Barnes2003a} proposed using an empirical relationship between stellar rotation, B$-$V color (as a proxy for stellar effective temperature), and age for main-sequence F-, G-, and K-dwarfs. This linear (in log-space) formalism for stellar rotation period, consisting of separable age and B$-$V color terms, represents the morphology and evolution of interface or "I-sequence" stars in color-period diagrams. These stars, thought to be rotating as solid bodies like the Sun, populate the upper-envelope of stellar rotation distributions. B03 provides an initial empirical calibration of this gyrochronology relationship using the few existing rotation period datasets for open clusters with well determined ages at that time. Subsequent large-scale photometric time-series surveys have led to updated calibrations using larger number of open clusters and rotation period measurements. Here, we also include in our analysis the I-sequence relationships calibrated in \citet[B07;][]{Barnes2007}, \citet[MH08;][]{Mamajek2008}, and \citet[MMS09;][]{Meibom2009}. We note that we use an updated age-dependence term given by \citet{James2010} for the MMS09 relationship.

\begin{figure*}
 \centering
 \includegraphics[scale=0.5,angle=0]{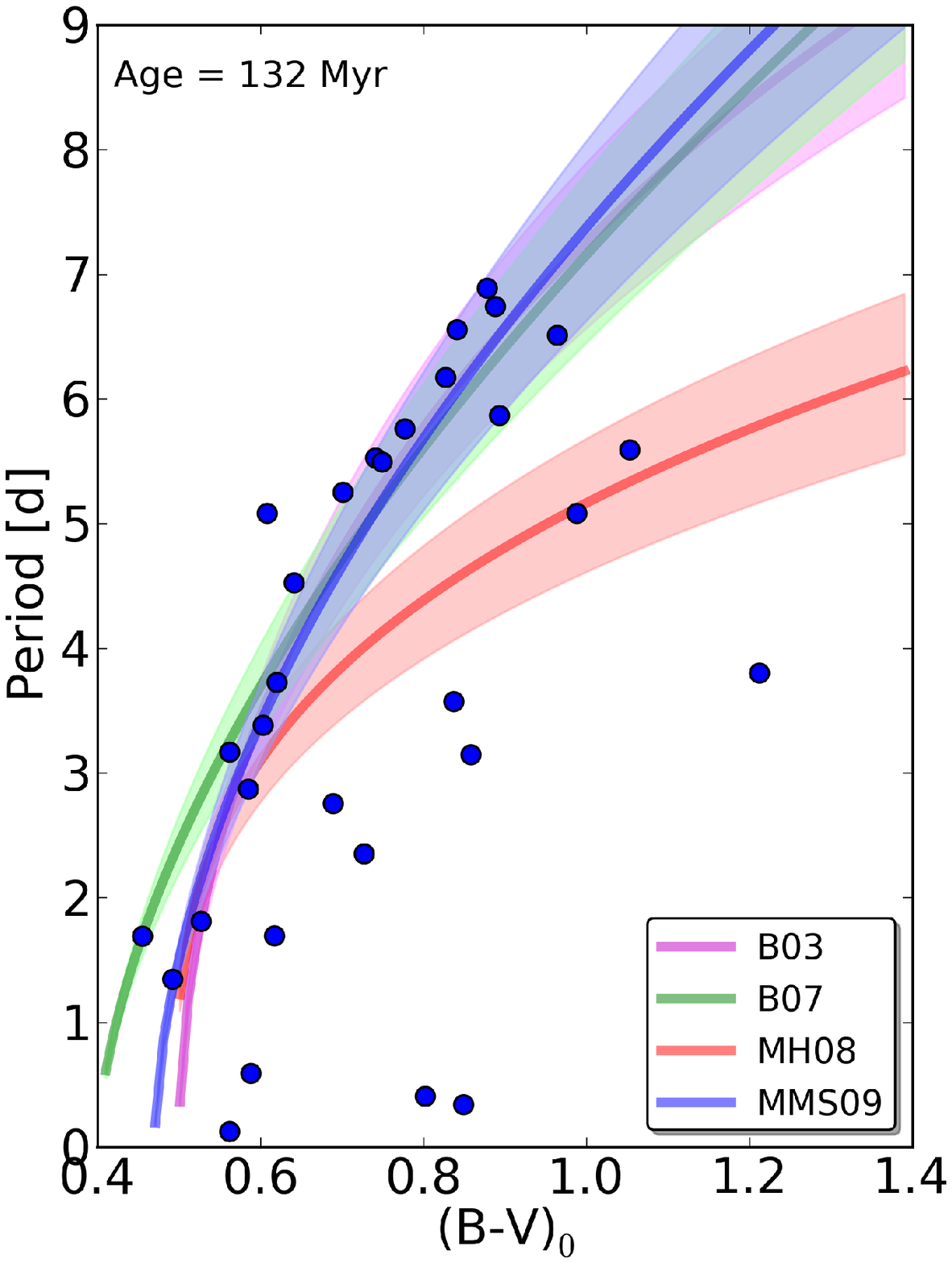}
 \includegraphics[scale=0.5,angle=0]{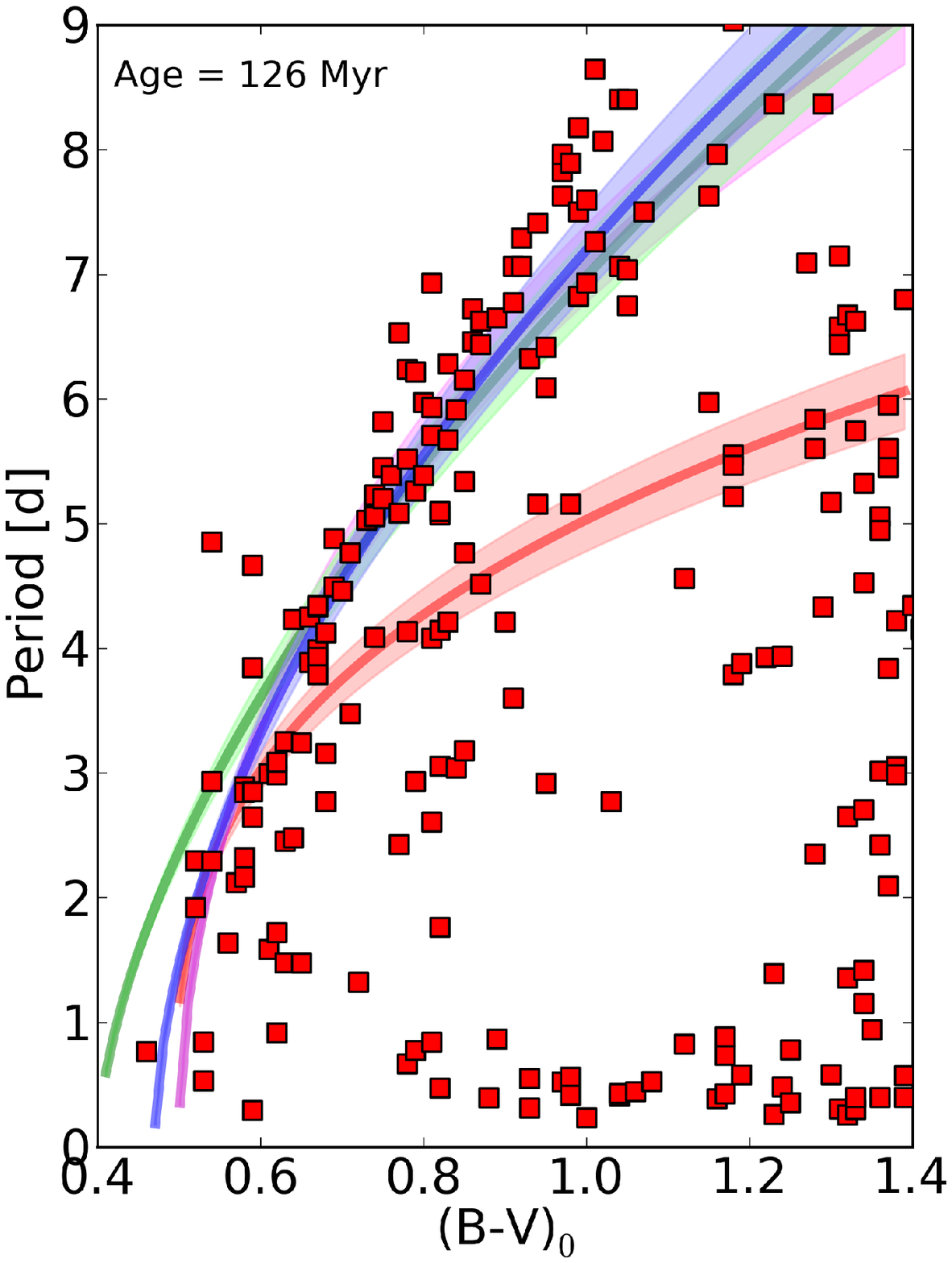}
 \caption{
   \label{fig.LDBIseq}
   Color-period diagram for Blanco~1 (left) and the Pleiades (right) with predicted I-sequence rotation isochrones at each cluster's respective LDB age (132$\pm$24 and 126$\pm$11 Myr). Shaded regions indicate the possible range in the predicted I-sequence location accounting for the uncertainty in the LDB ages for the clusters. Models used are -- B03: \citet{Barnes2003a}, B07: \citet{Barnes2007}, MH08: \citet{Mamajek2008}, and MMS09: \citet{Meibom2009}.
   }
\end{figure*}

In the left-hand plot of Fig.~\ref{fig.LDBIseq}, we compare the four gyrochronology relationships listed above (B03, B07, MH08, and MMS09) to the rotation period distribution of Blanco~1 while fixing the cluster age to that derived using its LDB location \citep{Cargile2010b,Burke2004}. The B03, B07, and MMS09 relationships appear to fit the upper-envelope morphology of the color-period distribution of Blanco~1 quite well, within their 3-$\sigma$ errors on period, while the MH08 relation fails to match the observational data redward of B$-$V$\sim$0.6 mag. We note the significant difference in the functional form of MH08 relationship (i.e., a flatter color-dependence) compared to the other gyrochronology relationships. Therefore, simply adjusting the age of Blanco~1 would not improve the overall match for the MH08. For example, by assuming Blanco~1 an older age would improve the agreement for MH08 for the redder stars, however, it would in turn cause the model to over-predict the rotation periods for the bluer cluster members. 

In the right panel of Fig.~\ref{fig.LDBIseq}, we plot the gyrochronology relationships with the color-period distribution for the Pleiades open cluster, and note that the above-mentioned Blanco~1 conclusions are the same. Namely, the B03, B07 and MMS09 models all appear to follow the morphology of the cluster's upper envelope of its color-period distribution, and again shows that the MH08 model falls considerably below the observed rotation periods for the reddest stars. We note however that all three of the B03, B07, and MMS09 models fail to match the upper envelope of the slow rotators for stars with B$-$V $>$ 0.80.
\subsection{Deriving a Rotation-Age for Blanco~1}
Employing the four gyrochronology I-sequence relationships (B03, B07, MH08, and MMS09), we are able to infer an age by determining the most probable fit to the upper-envelope of the Blanco~1 color-period distribution. We also include the rotation period distribution for the Pleiades cluster in this analysis. In order to use these relationships as reliable chronometers, one must determine the best way to identify the stars that make up the cluster's I-sequence. This involves isolating these stars from more rapidly rotating gap- and C-sequence stars, as well as accounting for the natural width present in the measured rotation rates of I-sequence stars due to the effects of differential rotation in stars, possible age spread amongst the stars in the cluster, period measurement errors, and any residual signature of the cluster's natal angular momentum distribution.
\subsubsection{I-sequence: Percentile-Rank Fitting}\label{subsubsec.PRF}
Our first approach to modeling the I-sequence in Blanco~1 and the Pleiades is by defining the upper-envelopes of the color-period distributions using percentile ranking in color bins. Within each color bin we defined the rotation period upper-envelope by calculating the period corresponding to the 75th percentile of the values in that bin. We define the uncertainty in this 75th percentile period by taking the difference between the 75th percentile and the bin's median divided by the square root of the number of data points in the bin. The 75th percentile was chosen as a more robust measure of the distribution's upper-envelope as compared to other estimates, e.g., taking the upper-most data point in each bin, which are biased to outliers clearly falling above the main distribution of points, as well as not accounting for the natural spread in the I-sequence.

We take a Bayesian parameter estimation approach to infer the I-sequence age posterior PDF for the four I-sequence models (B03, B07, MH08, MMS09) using an affine-invariant Markov-Chain Monte Carlo sampler \citep[{\em emcee}, ][]{Foreman-Mackey2013} to explore parameter space. As stated before, we bin the Blanco~1 periods by B$-$V color to derive the 75th percentile rankings. In order to account for any significant uncertainty due to the specific bin placement and/or size, we include the number of bins used and the bin width as nuisance parameters in our MCMC modeling. The most probable solutions for both Blanco~1 and the Pleiades resulted in binning the data with an average of $\sim$6 bins with widths of $\sim$0.15 mag in B$-$V. We also do not include any rotation period data that falls outside the B$-$V limits used in the original calibration of the individual I-sequence models.

\begin{figure*}
 \centering
 \includegraphics[scale=0.45,angle=0]{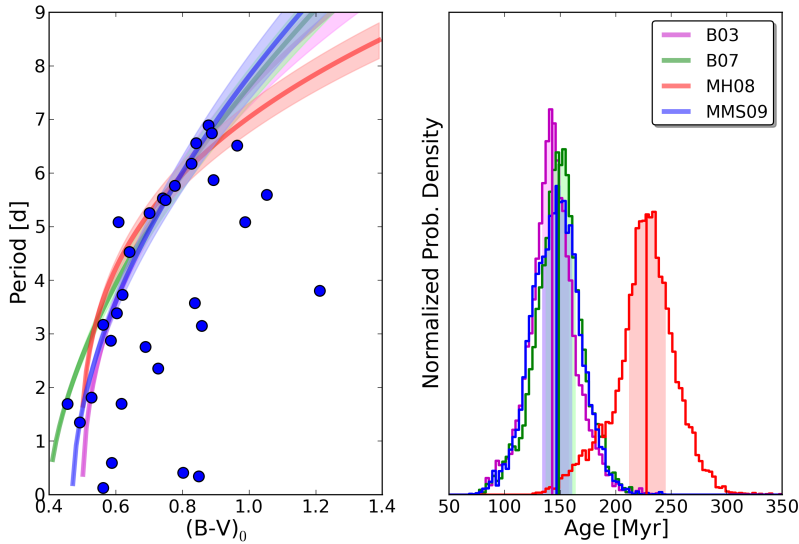} 
 \caption{
   \label{fig.gyrofit_B1}
   Color-period diagram for Blanco~1 stars (left) and I-sequence models at ages inferred from the PDFs based on the MCMC analysis (right). The shaded regions on plot indicate the range of the rotation periods based on the age-uncertainties as determined from the PDF interquartile ranges.
   }
\end{figure*} 

\begin{figure*}
 \centering
 \includegraphics[scale=0.45,angle=0]{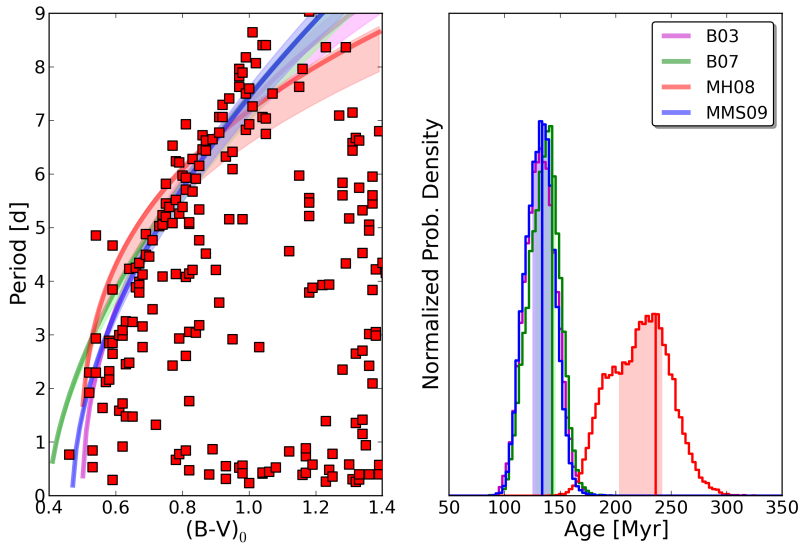} 
 \caption{
   \label{fig.gyrofit_Ple}
   Similar to Fig.~\ref{fig.gyrofit_B1}, but results from analysis of the Pleiades rotation period datasets (red data points).
   }
\end{figure*} 

Our MCMC chain consisted of 200 {\em walkers} and 500 steps (after a 100 step burn-in period), thereby, we obtained 100,000 independent model evaluations for each of the four I-sequence models. We determine the most probable I-sequence age of Blanco~1 and the Pleiades for each model based on the median value of the PDF, with uncertainties determined by the interquartile range. In Figs. \ref{fig.gyrofit_B1} and \ref{fig.gyrofit_Ple}, we plot the color-period diagrams for Blanco~1 and the Pleiades with the most probable I-sequence models, and the inferred rotation-age PDFs for each cluster. The I-sequence ages and uncertainties for each model are given in Table \ref{tab.gyrofit}.
\begin{deluxetable}{l c c c c c c c c}
\tablecolumns{5}
\tablewidth{0pc}
\tablecaption{I-Sequence Gyrochronology Ages for Blanco~1 \& the Pleiades \label{tab.gyrofit}}
\tablehead{
  \colhead{} & \multicolumn{2}{c}{Percentile-Rank} & \multicolumn{2}{c}{KDE} \\
  \colhead{} & \colhead{Age} & \colhead{B.F.\tablenotemark{a}} & \colhead{Age} & \colhead{B.F.\tablenotemark{a}} \\
  \colhead{Model} & \colhead{(Myr)} & \colhead{} & \colhead{(Myr)} & \colhead{}
}
\startdata
\cutinhead{Blanco~1}
B03    & 143$^{+12}_{-12}$ &  0.64    & 141$^{+8}_{-15}$  & 0.61 \\
B07    & 155$^{+6}_{-19}$  &  0.74    & 148$^{+13}_{-33}$ & 0.67 \\
MH08   & 236$^{+7}_{-24}$  &  \nodata & 224$^{+28}_{-106}$ & \nodata \\
MMS09  & 149$^{+12}_{-18}$ &  0.81    & 147$^{+4}_{-6}$   & 0.60 \\
\cutinhead{Pleiades}
B03    & 134$^{+8}_{-11}$ &  0.54    & 129$^{+8}_{-8}$   & 0.58 \\
B07    & 143$^{+2}_{-17}$ &  0.57    & 136$^{+9}_{-9}$   & 0.60 \\
MH08   & 236$^{+5}_{-34}$ &  \nodata & 198$^{+17}_{-74}$ & \nodata \\
MMS09  & 134$^{+7}_{-11}$ &  0.49    & 134$^{+5}_{-7}$   & 0.51 \\
\enddata
\tablenotetext{a}{The Bayes factor for each model as compared to MH08: B.F.$=$P(data$\vert$MH08)/P(data$\vert$B03,B07,MMS09). A B.F. $< 1$ suggests the respective models are more preferred to the MH08 model.}
\tablerefs{ {\footnotesize B03: \citet{Barnes2003a}, B07: \citet{Barnes2007}, MH08: \citet{Mamajek2008}, MMS09: \citet{Meibom2009}} }
\end{deluxetable}
For both clusters, the B03, B07, and MMS09 models have very comparable PDFs with approximately a common median and relative heights between the distributions. On the other hand, there is a strikingly dissimilarity between these models and the distribution from the MH08 model. Quantitatively, we can also see this based on their Bayes factors (i.e., the ratio of model posterior probabilities given the data). B03, B07, and MMS09 all have Bayes factors less than 1 when compared to MH08, allowing us to infer that these models are more favored to the MH08 model considering the Blanco~1 and Pleiades rotation period data. The Bayes factor is similar for B03, B07, and MMS09, suggesting that these models are nearly equally favored over the MH08 model. Keeping this in mind, we nevertheless still like to derive a single I-sequence age for Blanco~1 and the Pleiades for comparison purposes. Since the B03, B07, and MMS09 models provide comparable fits to the cluster upper-envelopes, we combine their PDFs into a joint probability density function and use the median of this distribution to derive an I-sequence age of 147$^{+13}_{-14}$($^{+20}_{-20}$) Myr for Blanco~1 and 134$^{+9}_{-10}$($^{+13}_{-14}$) Myr for the Pleiades. The quoted errors are based on the PDFs interquartile range, and in parentheses the inner 68\% of the PDF ($\sim$1-$\sigma$ errors assuming the PDFs are normal distributions). Redetermining these ages while including the MH08 model only results in a shift of $\sim$1 Myr older with a $\sim$10\% increase in the uncertainty. 

We also note the B03, B07, and MMS09 I-sequence models predicts gyrochronology ages that are consistent with the expected ages for both Blanco~1 and the Pleiades clusters as determined using the lithium depletion boundary technique (132$\pm$24 and 126$\pm$11 Myr for Blanco~1 and the Pleiades, respectively). Again, we see MH08 as an outlier in this sense; its predicted age for Blanco~1 and the Pleiades falls well beyond the 1-$\sigma$ LDB age-uncertainties for both clusters.
\subsubsection{I-sequence: Kernel Estimated Probability Density}
\begin{figure*}
 \centering
 \includegraphics[scale=0.45,angle=0]{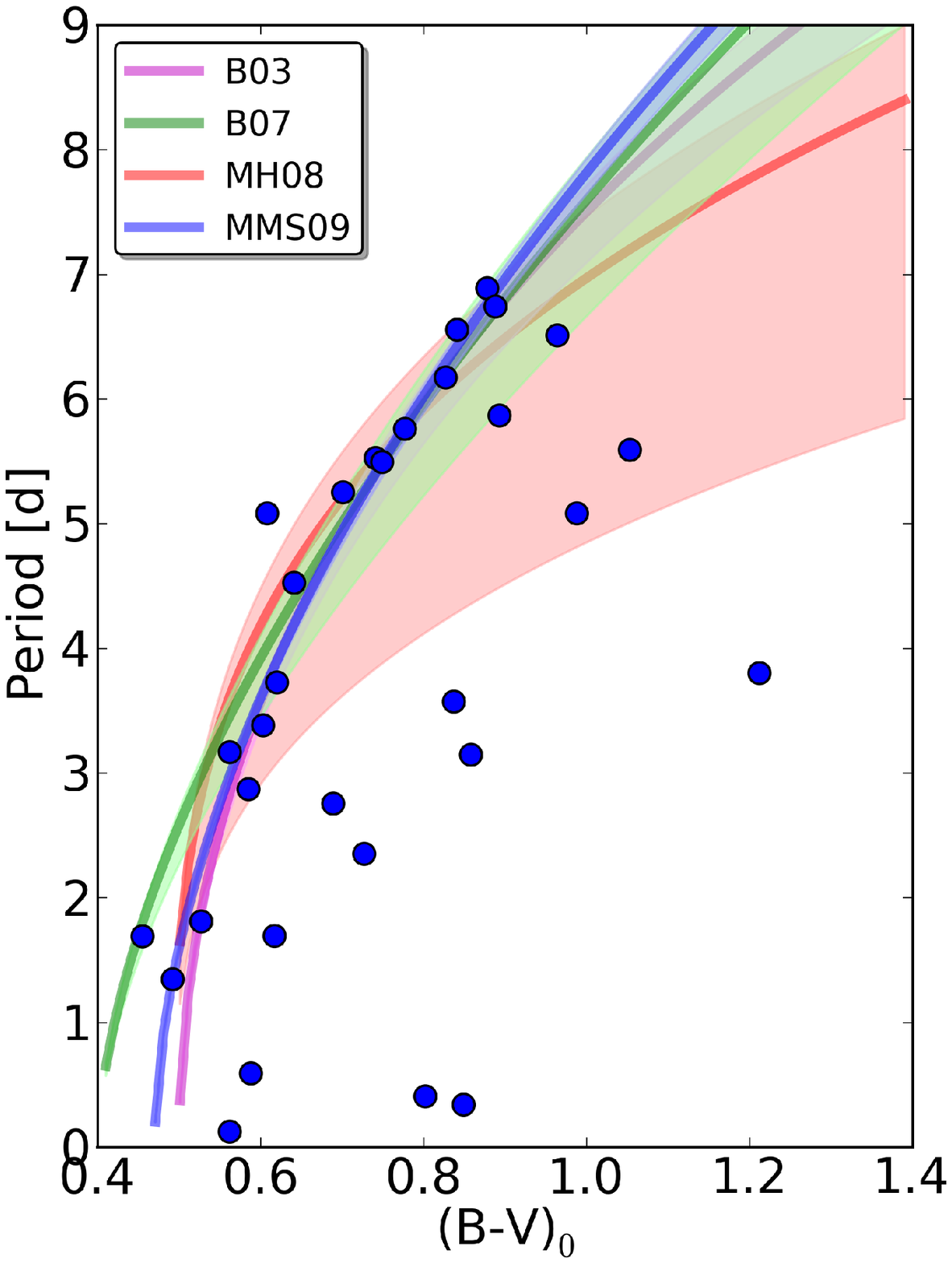} 
 \includegraphics[scale=0.35,angle=0]{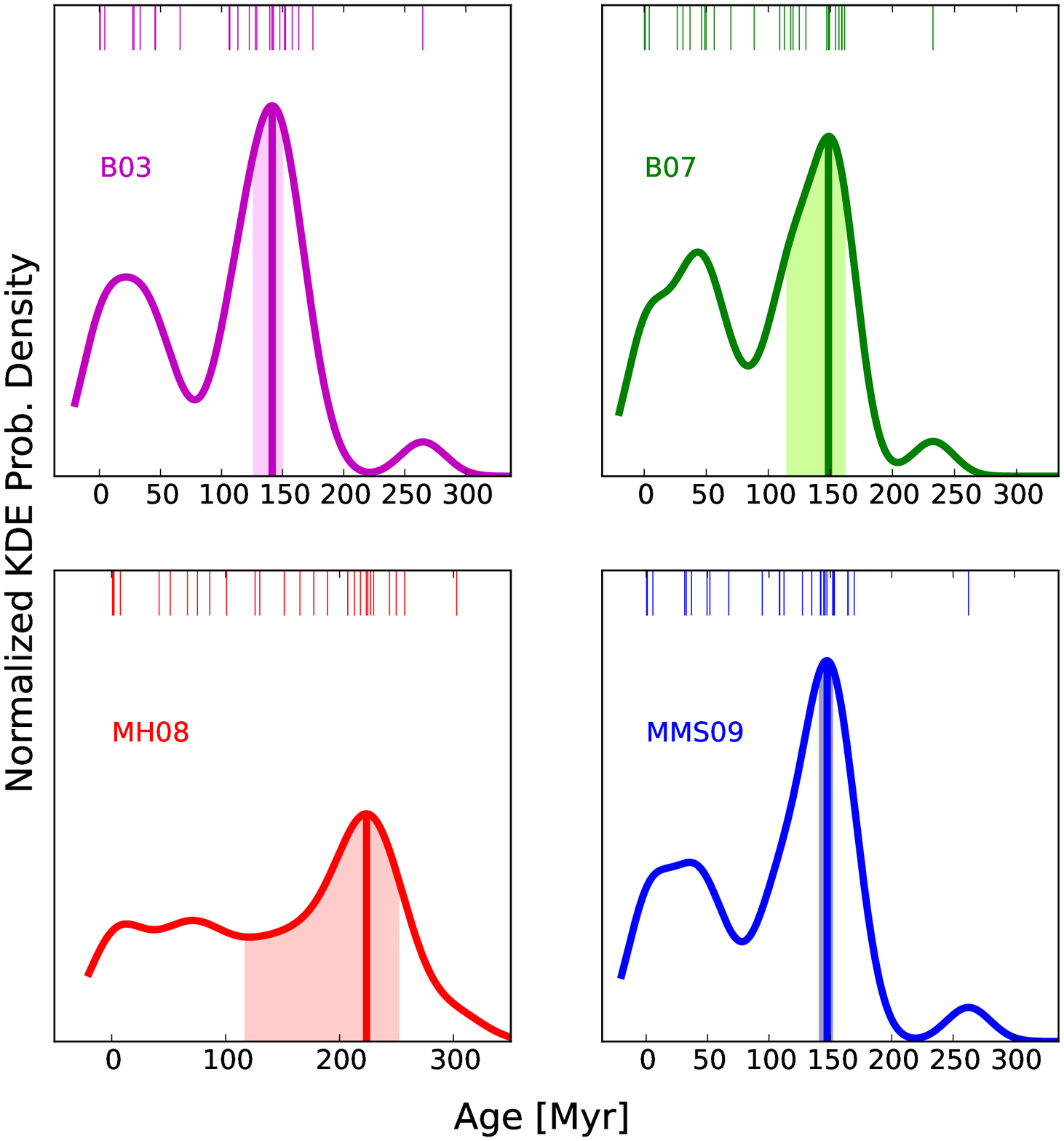} 
 \caption{
   \label{fig.B1_KDE}  
     Color-period diagram for Blanco~1 stars (left) and I-sequence models at ages inferred from their individual KDE probability density functions (right). The PDF plots have the same vertical linear scale for direct comparison. The shaded regions indicate the range of the rotation rates based on the age-uncertainties for the individual models as determined from the interquartile range resulting from a bootstrap error analysis. The rug-plot on the top axis of the KDE PDF plots indicate the individual age measurements. 
   }
\end{figure*}

\begin{figure*}
 \centering
 \includegraphics[scale=0.45,angle=0]{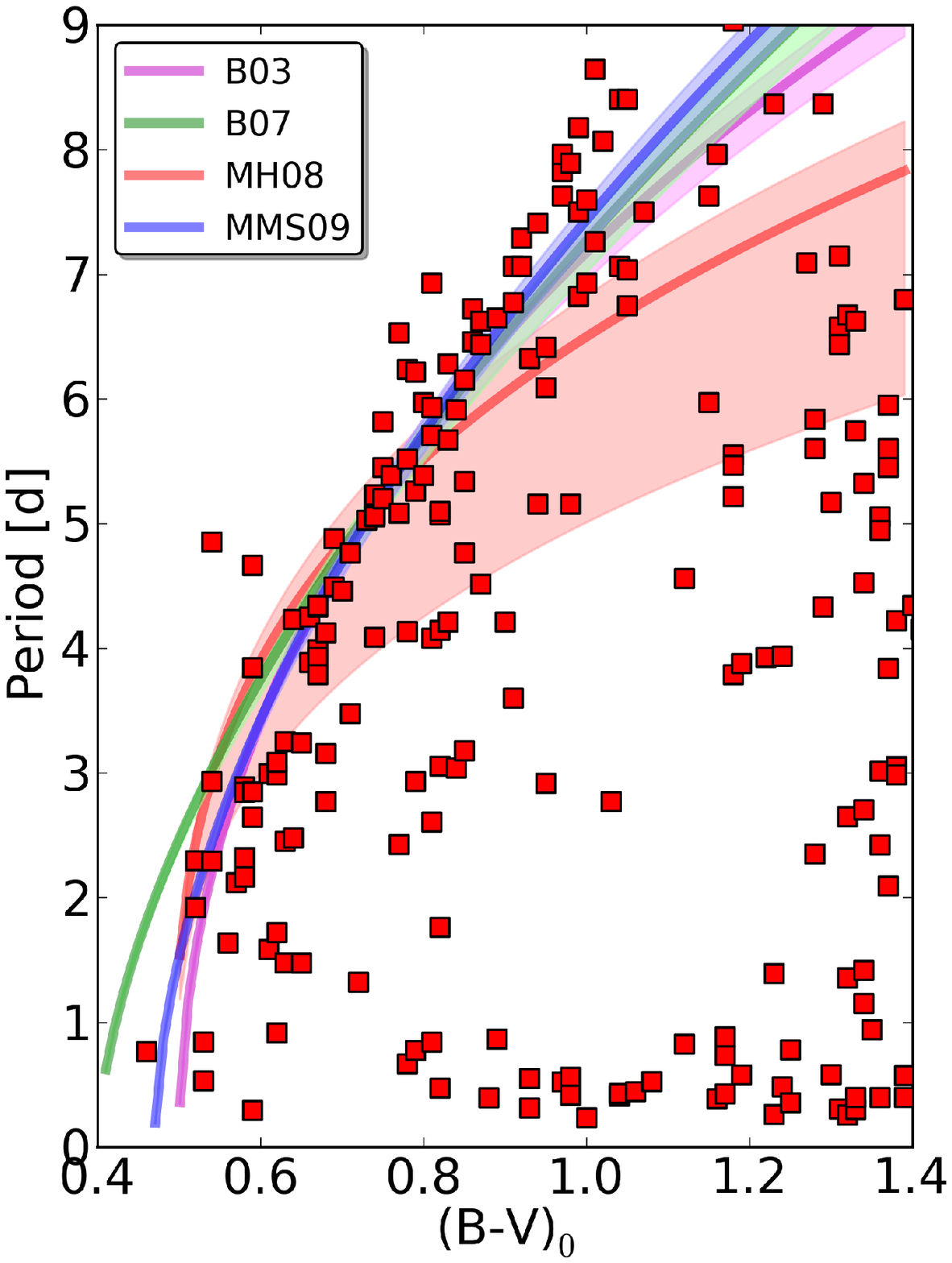} 
 \includegraphics[scale=0.35,angle=0]{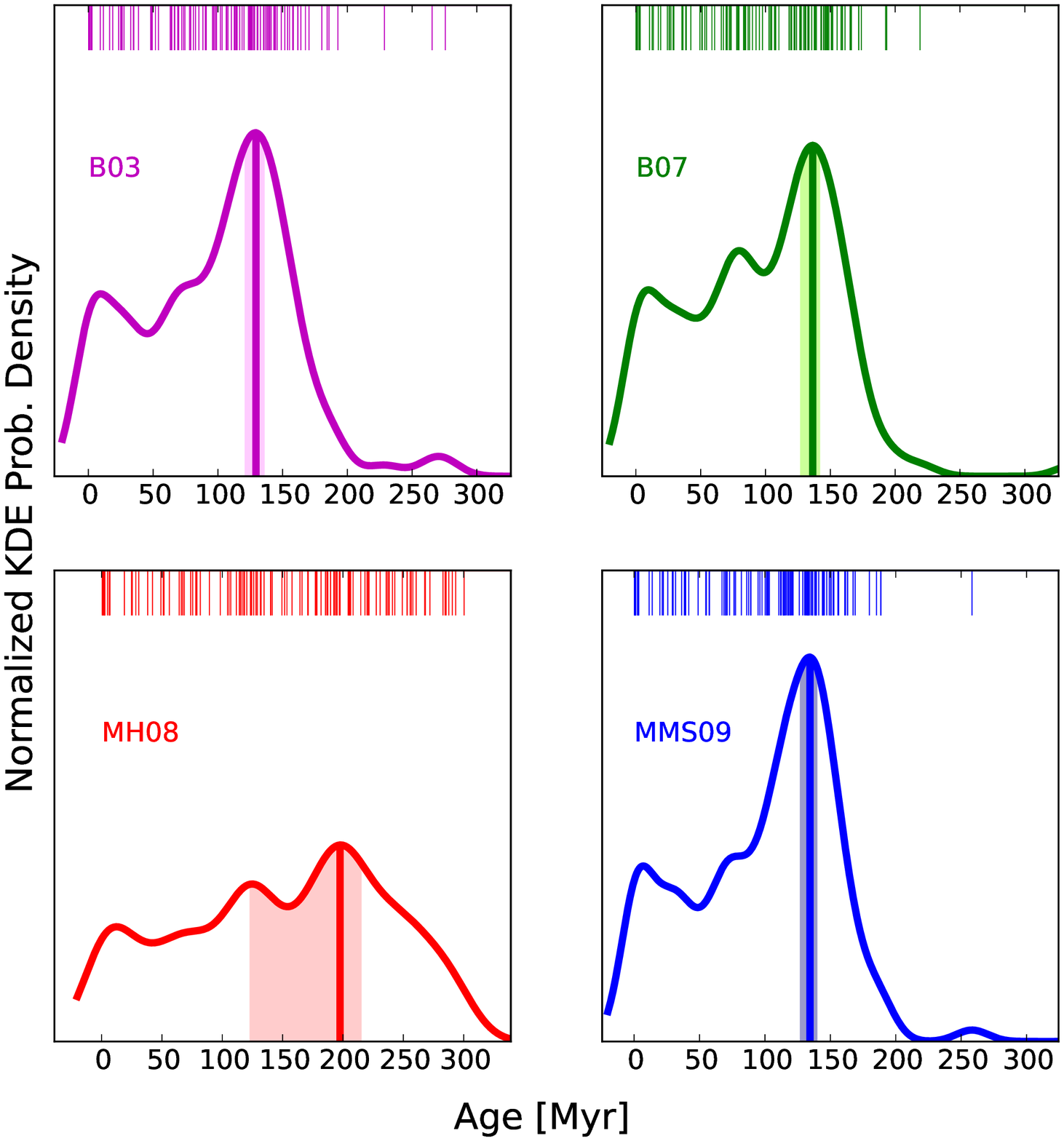} 
 \caption{
   \label{fig.Ple_KDE}  
   Similar to Fig.~\ref{fig.gyrofit_B1}, but results from analysis of the Pleiades rotation period datasets (red data points).
   }
\end{figure*}

In order to verify this first age-dating approach, we employ a second method to estimate the most probable gyrochronology I-sequence age for Blanco~1 and the Pleiades. We employ a KDE algorithm as a non-parametric way to estimate the PDF for a set data points. In our implementation, we use an I-sequence model to derive individual predicted ages for each cluster star using their measured rotation period and $B-V$ colors. We then compute the joint PDF for all of these ages using KDE with a univariate normal kernel. As with all KDE-based derivation of probability density functions, the most important parameter to define is the kernel bandwidth. Here, we determine this parameter using least squares cross-verification producing a width of $\sim$20 Myr, comparable to the expect error for individual stellar ages at $\sim$130 Myr using I-sequence gyrochronometry \citep{Barnes2007}. In Figs. \ref{fig.B1_KDE} and \ref{fig.Ple_KDE}, we show the resulting KDE PDF and inferred I-sequence models displayed on the color-period diagram for Blanco~1 and the Pleiades. We determine the most probable I-sequence gyrochronology age for Blanco~1 and the Pleiades by identifying the age with the maximum probability in the joint PDF for each model, and determine the uncertainty on these ages by calculating the interquartile range of 10,000 bootstrap samples with replacement (Table \ref{tab.gyrofit}).

In this analysis, we are not classifying stars as belonging to the I-sequence and/or defining an upper-envelope of the color-period diagram. Here, we are assuming that there exists a sequence of cluster stars which are best represented with an I-sequence relationship given by one of the four models we are considering here (B03, B07, MH08, and MMS09). If this assumption is true, then these stars will contribute significantly to a localized region of the probability density function at a specific age, thus resulting in a tall peak in the joint PDF. Therefore, the overall tallest peak in the joint PDF is defined as the most probable I-sequence age for the cluster. As stated in \citet{Barnes2003a}, I-sequence rotation models do not accurately represent the distribution of fast rotators (gap- or C-sequence stars) in open clusters. As a consequence, using I-sequence models to age-date these stars will result in systematically younger ages. For Blanco~1 and the Pleiades, the contribution from these C- and gap-sequence stars to their respective joint PDFs (see Figs. \ref{fig.B1_KDE} and \ref{fig.Ple_KDE}) is seen as a lower probability tail in the distributions at younger ages.

We find the B03, B07, and MMS09 I-sequence relationships produce similar probability densities for both clusters with their largest KDE peaks found approximately at the same age and near a common height. The KDE for the MH08 I-sequence model significantly disagrees with those produced by the other models. For both Blanco~1 and the Pleiades, MH08 has a lower overall KDE, suggesting a larger variation in the predicted I-sequence ages from this model, and the highest peak is found at a larger age for both clusters. As with our first method, the Bayes factor also quantitatively shows this trend, with B03, B07, and MMS09 models having a Bayes factor $<$ 1 compared to MH08 for both Blanco~1 and the Pleiades. We derive a single I-sequence age using our KDE analysis by computing a joint probability density function based on the B03, B07, and MMS09 KDE PDFs, disregarding the MH08 KDE due to it being a clear outlier. The resulting joint PDFs have median ages for Blanco~1 and the Pleiades of 145$^{+4}_{-102}$ and 133$^{+5}_{-81}$ Myr.

We note the significantly asymmetric uncertainties in these ages (determined from the interquartile range about the most probable age) is resulting from this contamination by rapid rotators, and therefore, accurately accounts for the difficulty in categorizing stars as being on the I-sequence. To better estimate the effect of these contaminants on our KDE-based I-sequence ages, we perform a least-squares minimization modeling each cluster's joint KDE PDF with two convolved Gaussian functions assuming it is a product of normal distributions for I-sequence stars and rapid rotators. For each cluster, the resulting best-fit model is composed of a narrow Gaussian near the expected cluster's age, and a broader distribution centered at $\sim$20 Myr. After subtracting the broader distribution from the joint KDE PDFs, we determine new I-sequence ages of 143$^{+19}_{-18}$($^{+28}_{-27}$) and 133$^{+17}_{-16}$($^{+25}_{-25}$) Myr for Blanco~1 and the Pleiades, respectively, based on the median of each cluster's corrected PDF. The errors are again based on the interquartile range, and in parentheses the inner 68\% of the PDF ($\sim$1-$\sigma$ errors assuming the PDFs are normal distributions).

These KDE-based I-sequence rotation-ages for Blanco~1 and the Pleiades are very much in agreement with the results found using the percentile-based technique in \S \ref{subsubsec.PRF}, providing evidence for the robustness of our I-sequence age determinations. And again, comparing our KDE-based I-sequence ages with previously determined LDB ages, we see this method also results in gyrochronology ages for Blanco~1 and the Pleiades that are consistent (within error) of their expected ages from the alternate distance-dependent age determination method.
\section{Evaluation of Stellar Spin-down Laws}
\subsection{Linear Spin-down Laws}
As stated in \S \ref{sec.Iseq}, the influence of mass and age on the spin-down of solar-type stars are assumed to be independent effects, resulting in a formalism that is a linear (in log-space) separable function describing the relationship between stellar rotation, mass (or color), and age \citep{Barnes2007,Barnes2010a}. The basic function used to map this relationship can be divided into a mass-dependent polynomial and an age-dependent power-law term, e.g., P(color or T$_{\rm eff}$,t)$=$f(color or T$_{\rm eff}$)$\times$g(t) with g(t)$=$t$^{\rm n}$. A Skumanich-like spin-down is represented with n$= 0.5$. Other studies have further calibrated the n parameter using open clusters of known ages, and the solar datum -- B07: n$= 0.5189$, MH08: n$= 0.566$, and MMS09 with the recalibration from \citet{James2010}: n$= 0.5344$.

\begin{figure}
 \centering
 \includegraphics[scale=0.35,angle=0]{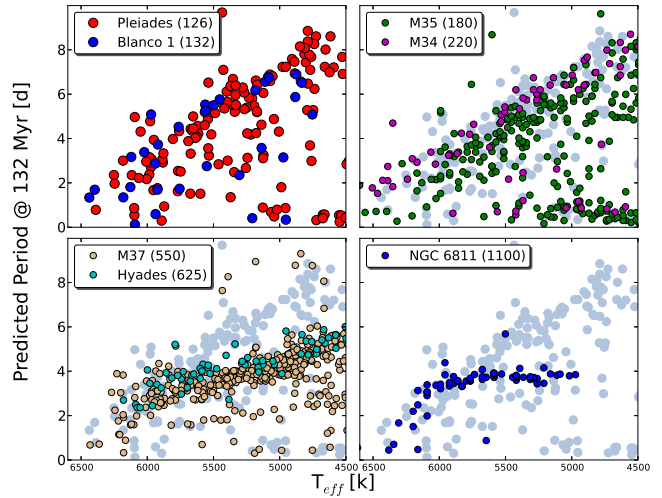} 
 \caption{
   \label{fig.linearspindown}  
   Predicted temperature-period diagram for various open clusters with their stars spun-up or -down to their expected rotation rate at the age of Blanco~1 (132 Myr) as inferred from the Skumanich spin-down law (i.e., $\times \sqrt{132/(Cluster Age)}$). The ages adopted for the individual clusters are: Blanco~1=132 Myr, Pleiades=126 Myr, M~35=180 Myr, M~34=220 Myr, M~37=550 Myr, Hyades=625 Myr, and NGC~6811=1.1 Gyr (also indicated in each individual legend). A common effective temperature scale was determined using the relationships given in \citet{Casagrande2010}. In the right and bottom panels, the gray points are Blanco~1 and Pleiades stars plotted for comparison.
	}
\end{figure}

MMS09 give evidence for different spin-down rates for G and K dwarfs by showing a Skumanich-like (n$=0.5$) spin-down law results in 625 Myr Hyades G dwarfs having periods in agreement with similar spectral type stars in the 150 Myr M~35 open cluster, however, K dwarfs appear to be rotating faster than what is predicted by this relationship. Here, we would like verify this finding using a set of rotation periods from a larger number of open clusters. Our sample consists of rotation periods for the following clusters with given ages: our new rotation periods for Blanco~1 (132 Myr), the HATnet periods for the Pleiades (126 Myr), M~35 (180 Myr) and M~34 (220 Myr) from \citet{Meibom2009,Meibom2011}, and M~37 (550 Myr) from \citet{Hartman2009}. We also include rotation periods from a new comprehensive catalog for the 625 Myr old Hyades (Kundert et al. in prep) and for the old (1.1 Gyr) open cluster NGC~6811 which was observed as part of the Kepler mission \citep{Meibom2011b}. In order to directly compare the rotation period distributions for each cluster, we converted the published photometric colors for individual stars, as provided in the source catalogs, to effective temperature based on the color-temperature relationships given in \citet{Casagrande2010}. 

The general method we follow is similar to that presented in MMS09, namely, we adjust the rotation periods for all clusters to their expected rotation rate at the age of Blanco~1. This requires us to multiply the rotation periods in the clusters using a scaling of $(Age_{Blanco~1}/Age_{cluster})^{\rm n}$, where n is the power-law coefficients listed above for the different models. In Fig \ref{fig.linearspindown}, we show the resulting rotation periods distributions as predicted by a Skumanich-like spin-down (n$=0.5$). Similar to MMS09, we too find higher-mass, solar-type stars in these various clusters agree well with the stars in Blanco~1. This confirms that in general these solar-type stars appear to be spinning down Skumanich-like from $\sim$100 Myr to 1 Gyr, rather unsurprising as the Skumanich-law itself was calibrated with solar-type stars over these same ages. 

However, we observe a break in stars cooler than the solar effective temperature ($\sim$5770 K), where it appears that stellar spin-down rates begin to deviate from a Skumanich-like rotation evolution. Stars with effective temperatures less than the Sun have systematically faster rotation rates compared to what is predicted by the Skumanich law for the age of Blanco~1. We note that the observed deviation occurs at a slightly higher effective temperature than what was originally suggested in MMS09 (these authors suggest the break from Skumanich is observed in cooler K dwarfs). Furthermore, we see this effect is enhanced in low-mass stars at ages $\gtrsim$ 1 Gyr, suggesting that the phenomenon continues beyond the age of the Hyades for low-mass solar-type stars with increasing disagreement from what is predicted by Skumanich-like spin-down. We have also tested other linear spin-down laws from B07, MH08, and MMS09, and due to the similarity in their age dependences to the Skumanich relationship (specifically, their value of the n exponent in the g(t) term), we find they do not improve the observed dependencies at lower mass.

The observation of less efficient, non-Skumanich-like spin-down for stars cooler than the Sun provides further empirical evidence supporting stellar angular momentum models that predict a more complex spin-evolution than a simple linear spin-down. In fact, B03 suggests that the observed I-sequence morphology seen in open clusters can be interpreted as the spin-down evolution of the full stellar interior (i.e., as a solid-body rotator). However, more recently \citep{Barnes2010a} show the I-sequence does not follow what is predicted from this simple assumption based on stellar structure models, instead finding a more consistent relationship between the mass-dependent convective turn-over timescale and observed I-sequence morphology. Also, recent theoretical studies have outlined other mass-dependent sources of angular momentum redistribution during solar-type and low-mass stellar evolution, including interior processes (gravity waves, core-envelope coupling, diffusion processes, etc.) and wind breaking formalisms \citep[e.g., ][]{Charbonnel2005,Irwin2009,Denissenkov2010,Spada2011,Matt2012}. In the following section we examine the rotation distributions of Blanco~1 and the Pleiades in the context of more complex, non-linear angular momentum models. Nevertheless, using a comparative analysis similar to Fig. \ref{fig.linearspindown} with additional data from high fidelity age-dated open clusters, in particular clusters older than $\sim$200 Myr, should allow recalibrating the empirical I-sequence model providing a more accurate formalism.
\subsection{Non-Linear Spin-down Laws}\label{subsec.NLSD}
Recently, several non-linear (i.e., the mass- and age-dependence on rotation period cannot be described as independent, separable functions) theoretical models have been put forward to explain the observed spin-down evolution of solar-type stars \citep{Barnes2010a,Barnes2010b,Reiners2012,Matt2012,Gallet2013}. These approaches use more comprehensive wind acceleration models that predict the amount of torque applied to different mass stars as they evolve, and include additional complexities such as saturation of angular momentum at high rotation rates \citep{Chaboyer1995}. A common feature in these models is to explain the distribution of rotation periods seen in young open clusters as a result of the distribution of initial rotation rates ($P_{0}$) observed in pre-main-sequence stars. These models predict that the stars ``forget" the signature of their initial rotation rate as they spin-down on the main-sequence at older ages. Therefore, young open clusters (ages $<$ $\sim$500 Myr) are valuable to test the initial stellar rotation distribution and this prediction from the angular momentum models.

\begin{figure}
 \centering
 \includegraphics[scale=0.4,angle=0]{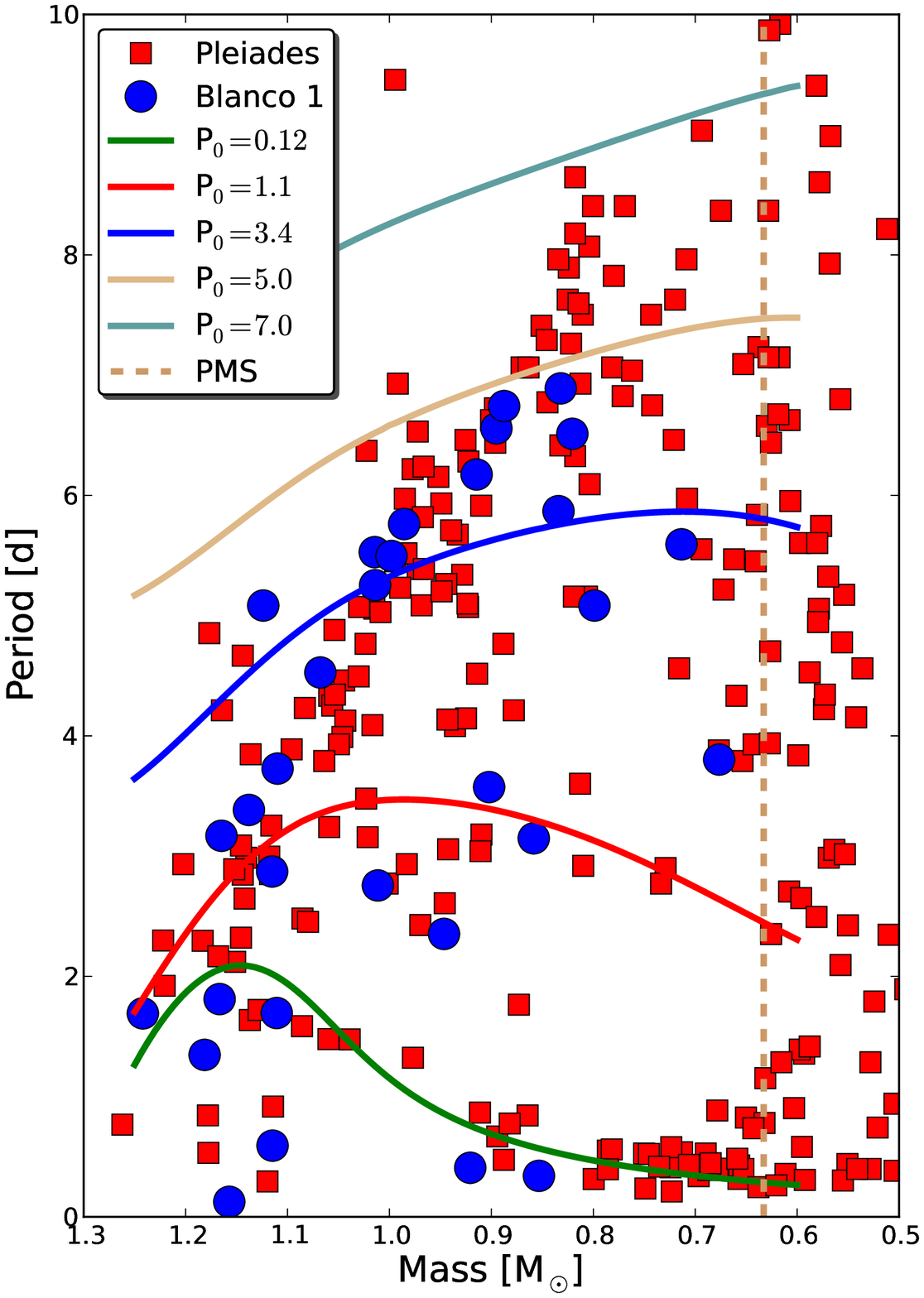} 
 \includegraphics[scale=0.4,angle=0]{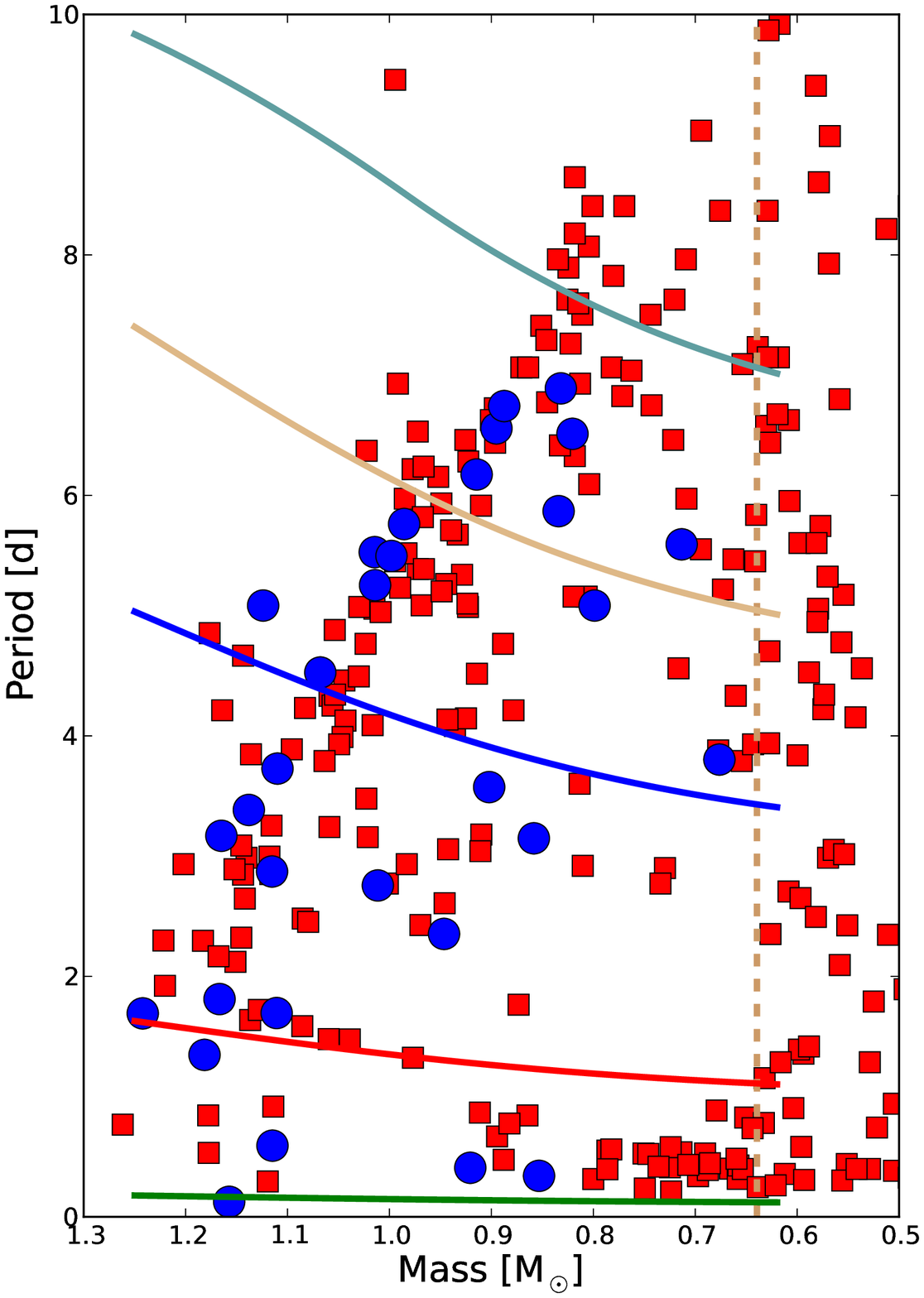} 
  \caption{
   \label{fig.angmom} 
 	Mass-Period diagrams for Blanco~1 (blue points) and the Pleiades (red points) stars with measured rotation periods. Also plotted are the loci of expected rotation rates for stars at 130 Myr with different initial rotation rates (P$_{0}$, see legend for specific values) as predicted from B10 (top) and RM12 (bottom) non-linear spin-down models. Stars still contracting on the pre-main-sequence are found to the right of the vertical dashed lines in both plots.
   }
\end{figure}

Using our measured rotation periods for Blanco~1 stars, coupled with the HATnet sample of Pleiades rotation periods, we can directly test these angular momentum evolution models. Here, we evaluate the predictions from \citet[B10; ][]{Barnes2010a} and \citet[RM12; ][]{Reiners2012}. We refer the reader to the detailed derivation of these rotation-isochrones given in B10 and RM12. The B10 and RM12 models predict rotation rates as a function of stellar mass, therefore, we convert the effective temperatures (calculated using \citealt{Casagrande2010} relationship between color and temperature) for each Blanco~1 and Pleiades star into mass using the Pisa stellar evolution models \citep{Tognelli2011} at 132 and 126 Myr for Blanco~1 and the Pleiades, respectively. A unique feature of the RM12 model is that its predicted angular momentum loss rate is radius dependent, therefore, we also use the PISA models to calculate radii for the Blanco~1 and Pleiades stars. In Fig.~\ref{fig.angmom}, we show the predicted rotation rates at 130 Myr for stars with a range of $P_{0}$ values. Each line in Fig.~\ref{fig.angmom} shows the predicted rotation rate of stars at 130 Myr given they reached the zero-age main-sequence with a certain $P_{0}$ value (indicated by the different colored lines). Observations of young stellar populations show that typically stars are rotating near breakup speed (for the Sun, $P_{break up} = 0.12$ d), but with a possible range up to $\sim$10 d \citep{Herbst2005}. Therefore, at any given stellar mass, the predicted range of rotation rates inferred by this observed distribution of $P_{0}$ is seen as the total vertical spread between the plotted lines. We have also indicated the pre-main-sequence ``turn-on'' mass as predicted by the PISA models. Stars more massive than this are on the main-sequence and have their angular momentum evolution dominated by spin-down due to magnetized winds. The angular momentum evolution of less massive stars is governed by a combination of spin-up due to their contracting radius, and spin-down due to stellar magnetized winds. RM12 includes the effect of increased rotation rate as stars contract, however, the B10 model does not and is therefore only applicable to stars on the main-sequence.

Although both models predict a spread in rotation periods similar to that observed due to a range of $P_{0}$ values, it is clear neither can completely describe the overall morphology of rotation periods in Blanco~1 and the Pleiades. Using these models to explain the observed period distribution would require a distinct mass dependence on $P_{0}$. Specifically, high-mass solar-type stars in these clusters would be predicted to have a smaller range of initial rotation rates compared to lower mass cluster stars, a trend not seen in other young open clusters. Moreover, these models assume $P_{0}$ is the rotation period at which the star begins to undergo spin-down governed by their respective angular momentum loss laws, and they assume the star has this period at the zero-age-main-sequence. An alternative explanations for the disagreement could be that the star experiences spin-down at a different rate than what is predicted by these wind models. Recently, \citet{Matt2012} showed that torque on a star due to a magnetized wind is a complex function of stellar properties, as well as magnetic field strength and topology, properties that are known to evolve in young, rapidly rotating stars \citep{Gregory2012}.

Also, neither model is able to explain the apparent over-density of stars at all masses along the upper-envelope (I-sequence) of the distribution. In fact, RM12 also identifies this disagreement with their models and the upper-envelope of rotation distributions at $\sim$100 Myr, and speculates that this is due to the decoupling of the stellar radiative core and outer convective envelope in rapidly rotating solar-type stars. There is a slight improvement in the agreement with the upper-envelope morphology and B10 models for the highest mass stars ($>$1.1 M$_{\odot}$) in Blanco~1 and the Pleiades. Namely, the overall spread in predicted rotation rates at 130 Myr decreases for these stars and the predicted periods increase with decreasing mass. These effects are a result of the B10 model predicting that stars converge to an I-sequence distribution at a rate proportional to their global convective turn-over timescales, a parameter characterizing the depth and typical fluid velocities found in stellar convection zones \citep{Kim1996}. High-mass, solar-type stars are predicted to have short convective turn-over timescales due to their small convection zone depths, thus these stars are predicted by the B10 models to converge I-sequence distribution the earliest. In any case, it is apparent that there is a clear need for further investigation into why these models currently disagree with the observed rotation period distributions of Blanco~1 and the Pleiades.
\section{Summary and Conclusions}
We present the results from a time-series photometric survey of the Southern open cluster Blanco~1. The photometric light curve data for this study comes from the KELT-South telescope during its commissioning phase. Using a proper motion membership catalog for Blanco~1, we identified light curves for 94 high fidelity cluster members in the KELT-South dataset. These matched cluster stars were found to have V$\sim 8-15$ with a range of spectral types extending from early-A to late-K dwarfs. From these 94 light curves, we identified 35 objects with significant and reliable periodic variability in the KELT-South light curves dataset. The periods for these Blanco~1 stars were measured using a standard Lomb-Scargle periodogram technique with significance determined using FAP statistics. The majority of stars exhibiting periodic variability distinctly fall along the cluster sequence of Blanco~1 in both $B-V$ and $V-I_{c}$ CMDs. At high-mass end, we observe 2 late-A or early-F dwarfs showing variability potentially due to pulsations, however, further observations are required to confirm their nature. Apart from 2 photometric non-members, the other 31 stars in our sample are F-, G-, or K-dwarf cluster members, with periodic variability assumed to be from rotation induced light modulation due to stellar spots, with periods ranging from ultra-fast rotators with periods $\sim$ 0.1 days to slower-rotators with periods of $\sim$ 8 days.

We find a very similar morphology in the rotation period distributions for Blanco~1 and the coeval Pleiades cluster, even though these two clusters most likely have significantly different formation and evolution histories. This suggests that the individual stellar rotation properties and angular momentum evolution must not significantly depend on the overall broader cluster properties, as well as provides some evidence that indicates that factors altering individual stellar rotation rates (e.g., tidally-locked close binary systems) must occur with similar frequency in both clusters, regardless of the clusters' past histories.

The rotation period distributions of Blanco~1 and the Pleiades allow us to test the accuracy of commonly used I-sequence empirical gyrochronology relationships by comparing different formalisms for the color-dependent relationship of stellar rotation and age. Fixing the I-sequence age of Blanco~1 and the Pleiades to those determined by the LDB technique (132 and 126 Myr, respectively), we observe that the overall morphology of the I-sequence in Blanco~1 and the Pleiades is best predicted by B03, B07, and MMS09 models. For both clusters, the MH08 I-sequence model does not generally follow the I-sequence morphology systematically under-predicting the periods for stars redder than B-V$\sim$0.65.

Using these I-sequence models, we employ two separate methods to infer the gyrochronology age of Blanco~1 and the Pleiades. Our first technique consists of a MCMC analysis to determine the most probable age for each model. Based on this analysis, we find I-sequence ages for Blanco~1 and the Pleiades of 147$^{+14}_{-16}$ and 134$^{+9}_{-10}$ Myr, respectively, and for both Blanco~1 and the Pleiades, the B03, B07, and MMS09 I-sequence models are more favorable compared to MH08 based on the available cluster rotation periods. In our second approach we use a KDE to determine the probability density function of the predicted age for all of the stars in Blanco~1 and the Pleiades using the four I-sequence models. We determine the most probable age for each cluster by determining the median of a combined KDE PDF, resulting in ages of 143$^{+19}_{-18}$ Myr for Blanco~1 and 133$^{+17}_{-16}$ Myr for the Pleiades. Using this technique, we again find the B03, B07, and MMS09 I-sequence models best represent the cluster rotation periods as compared to the MH08 model. Overall, the general consistency of the ages inferred using our two techniques gives evidence that they are accurate determinations of the clusters' gyrochronlogical ages based on the I-sequence model framework.

Our I-sequence ages for Blanco~1 and the Pleiades provide unique opportunities to compare the ages derived using gyrochronology and the LDB technique. Due to limitations in the application of both techniques, a comparison between these two stellar chronometers has not previously been undertaken. The ages derived for open clusters using these two techniques are nearly independent\footnote{There is a slight degeneracy in gyrochronology and LDB ages due to the empirical calibration of the I-sequence relationships using stellar evolution models also employed in the LDB technique. However, we predict this degeneracy to be small as the two stellar chronometers are being applied to vastly different groups of stars. Gyrochronology is only applicable to F, G, K dwarfs on the main-sequence, and LBD ages are derived using measurements from very low-mass pre-main-sequence cluster members.}. The LDB technique is not sensitive to the particular pre-main-sequence model used to determine the cluster age, but it is very dependent on systematic uncertainties in magnitude-to-luminosity conversion \citep[e.g., bolometric corrections, the assumed distance to the cluster, etc][]{Burke2004}. I-sequence gyrochronology ages are distance independent by nature, but the accuracy of the models is fundamentally tied to the fidelity of their empirical calibration. The consistency of the I-sequence and LDB ages for Blanco~1 and the Pleiades gives strong evidence for their overall chronometric accuracy.

In all of our various evaluations of the four I-sequence models, we consistently find a lack of agreement between MH08 model and rotation period distributions in the Pleiades and Blanco~1. As stated in Sec. \ref{sec.Iseq}, this appears to be due to the flatter color-dependence in the MH08 relationship between age and B$-$V compared to the other models. We note B03, B07, and MH08 all include the Pleiades in their original calibration, including using similar rotation period catalogs for the cluster. However, B03 uses a Pleiades age of 100 Myr, while B07 and MH08 use values (120 and 130 Myr) more similar to the cluster's LDB age of 126 Myr. Interestingly, using our larger sample of Pleiades rotation periods (the HATnet sample overall increase the number of Pleiads with known rotation periods by a factor of $\sim$5), we recover a cluster age that is more consistent with 130 Myr using the B03 and B07 models, but infer a much older ($\sim$ $+$100 Myr) age using MH08 model. Therefore, based on our analysis of Blanco~1 and the Pleiades we must caution against the use of the MH08 I-sequence model when deriving the gyrochronology age for stellar populations or individual stars with ages near $\sim$100 Myr. Further investigation is needed to determine if this disagreement is limited to zero-age-main-sequence stars, or whether this trend extends to older/younger stellar ages.

Using the rotation periods for Blanco~1, we test a basic spin-down model where the rotation rate of a star of a given mass/temperature is linearly dependent (in log-space) upon its age. Using a large sample of rotation periods from open clusters spanning ages from $\sim$130 Myr to 1.1 Gyr, we employ the Skumanich relationship (rotation $\sim 1/\sqrt{Age}$) to spin-up (or -down) the individual cluster stars to their rotation rate predicted at the age of Blanco~1. Comparing these adjusted periods to the rotation distribution of Blanco~1 we see stars with Teff $\geq$ Teff$_{\odot}$ are in agreement, thus suggesting that their rotation evolution is accurately represented using the Skumanich relationship. However, cooler stars (Teff $\lesssim$ 5700 K) show a distinct break from the Skumanich law, namely, they appear to be rotating faster at the age of Blanco~1 than that predicted by a $1/\sqrt{Age}$ law. We investigated this phenomenon using slightly adjusted values for the power-law dependence (e.g., n$=$0.52-0.57, as seen in B03, B07, MH08, and MMS09) without success in reproducing the rotation period data. The observed deviation suggests that there is a significant temperature/mass dependency in the spin-down and angular momentum loss rate for stars cooler than the Sun with ages from $\sim$100 Myr to 1 Gyr. This provides strong empirical evidence supporting more complex, mass-dependent spin-down laws that do not rely on simple assumptions (e.g., solid body rotation) that are required to explain linear, Skumanich-like rotation evolution.

Finally, our comparison of rotation period distributions of Blanco~1 and the Pleiades with stellar non-linear angular momentum evolution models of B10 and RM12 show significant disagreements between the predicted rotation rates from these models and the observed morphology of these open clusters. In light of our comparison of these models with rotation periods from stars in Blanco~1 and the Pleiades, these models can be used as qualitative guides to understand the overall rotation evolution of solar-type stars, and moreover encourage further theoretical development in order for them to be used to derive quantitative assessments of stellar rotation distributions including their use as tools for gyrochronology.

\acknowledgments
P.A.C. acknowledges the support from the National Science Foundation Astronomy and Astrophysics Research Grant AST-1109612. KELT-South received funding from the Vanderbilt Initiative in Data-Intensive Astrophysics (VIDA), FISK-Vanderbilt NSF PARRE grant AST-0849736, and the Vanderbilt International Office. Observational data reported in this publication are based on data obtained from the KELT-South telescope at the South African Astronomical Observatory (SAAO), Sutherland, South Africa. 
\clearpage
\appendix
\section{Phased Light Curves}
This appendix presents the LSP and phased KELT-South light curves for the 35 Blanco~1 stars with measured periods. For each star, we plot the periodogram on the left and its light curve phased to our measured periods. In the periodograms, we identify the stars using the KELT ID listed at the top of the plot, and the peak period is indicated on the periodograms by an arrow above the plot. Binned magnitudes are also plotted in the light curve plots, calculated in 10 equal size bins, however, the periodograms have been determined using the full light curve. 
\begin{figure}[!h]
 	\figurenum{14}
	\label{fig.phasedlightcurves}
	\includegraphics[scale=0.5,angle=0]{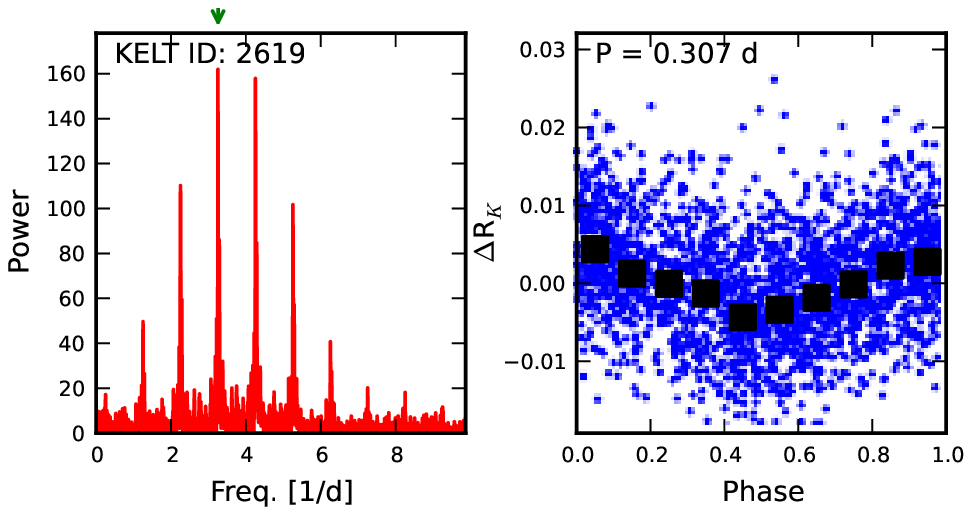}
	\includegraphics[scale=0.5,angle=0]{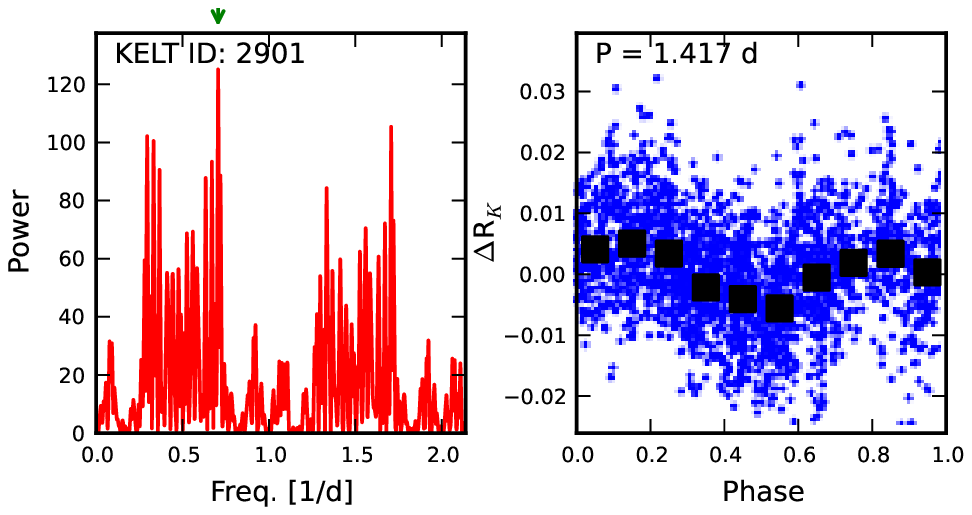}
	\includegraphics[scale=0.5,angle=0]{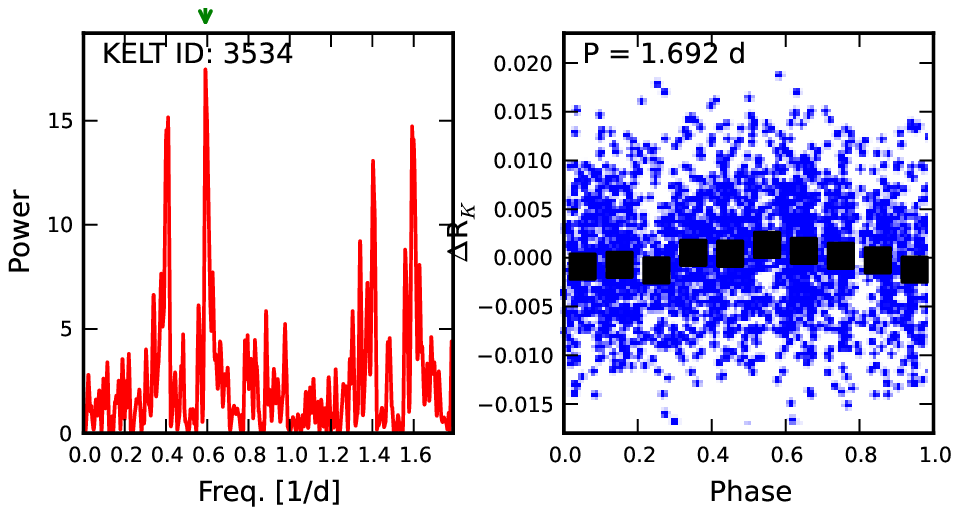}
	\includegraphics[scale=0.5,angle=0]{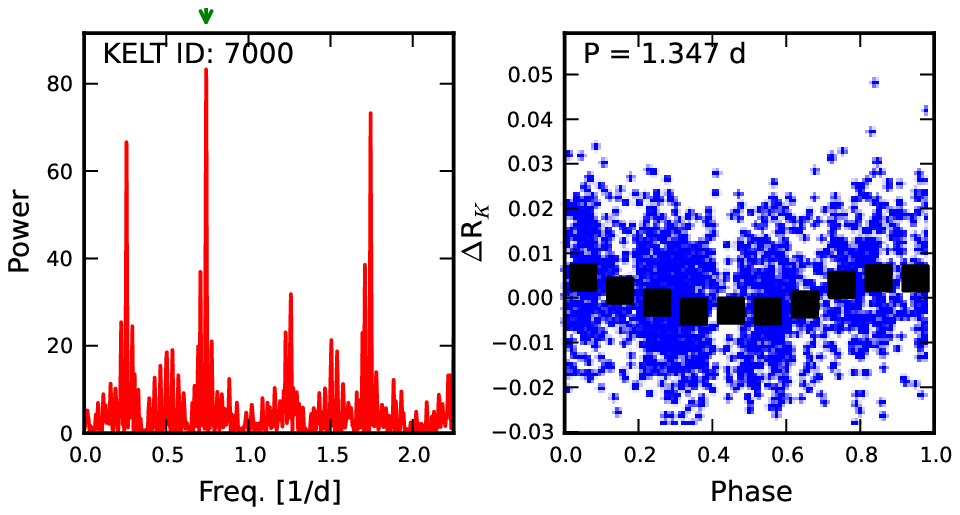}
	\includegraphics[scale=0.5,angle=0]{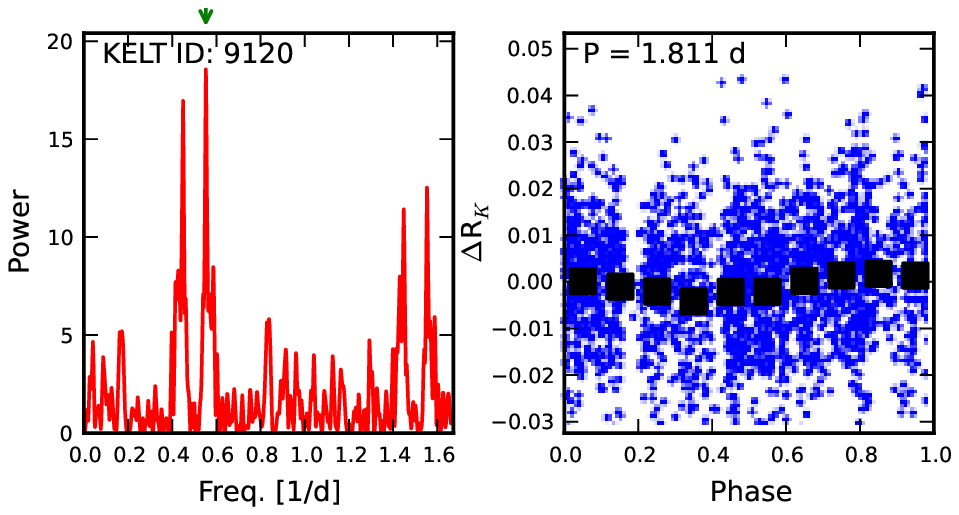}
	\includegraphics[scale=0.5,angle=0]{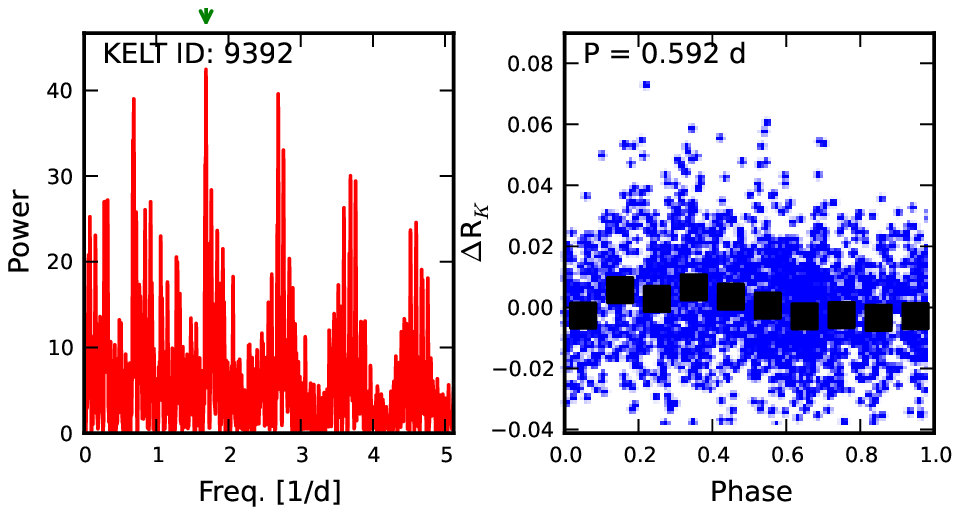}
	\includegraphics[scale=0.5,angle=0]{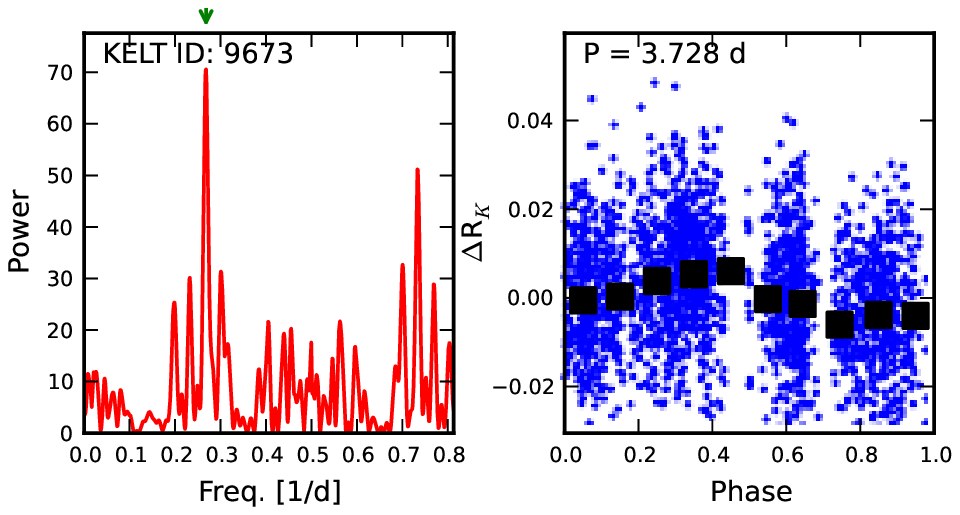}
	\includegraphics[scale=0.5,angle=0]{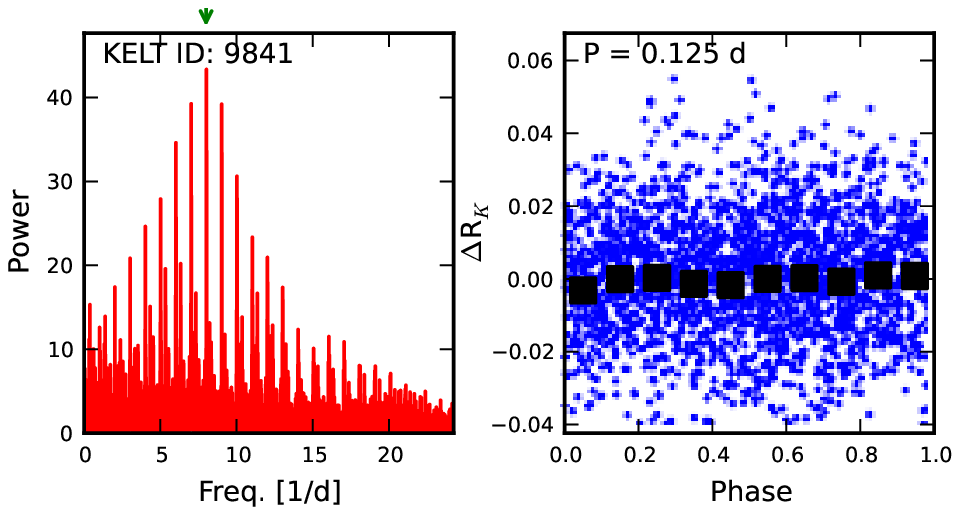}
	\includegraphics[scale=0.5,angle=0]{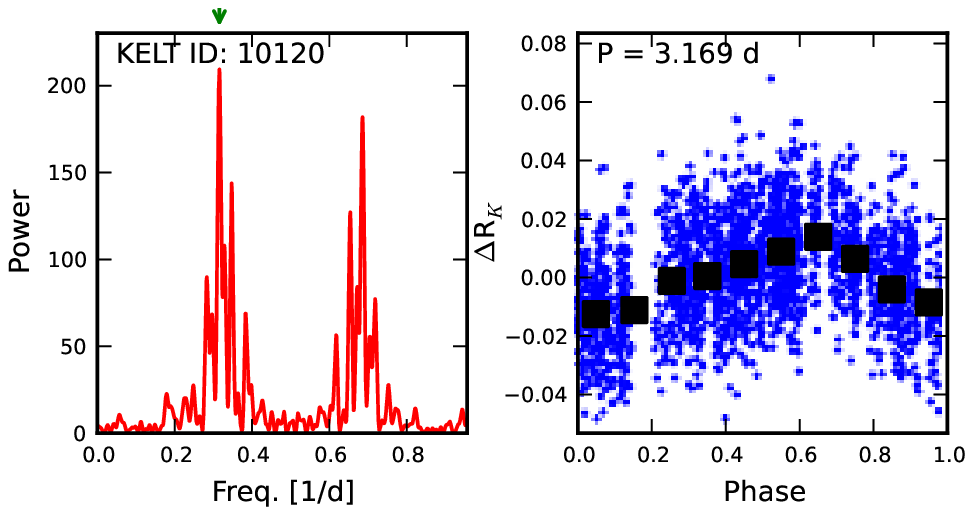}
	\includegraphics[scale=0.5,angle=0]{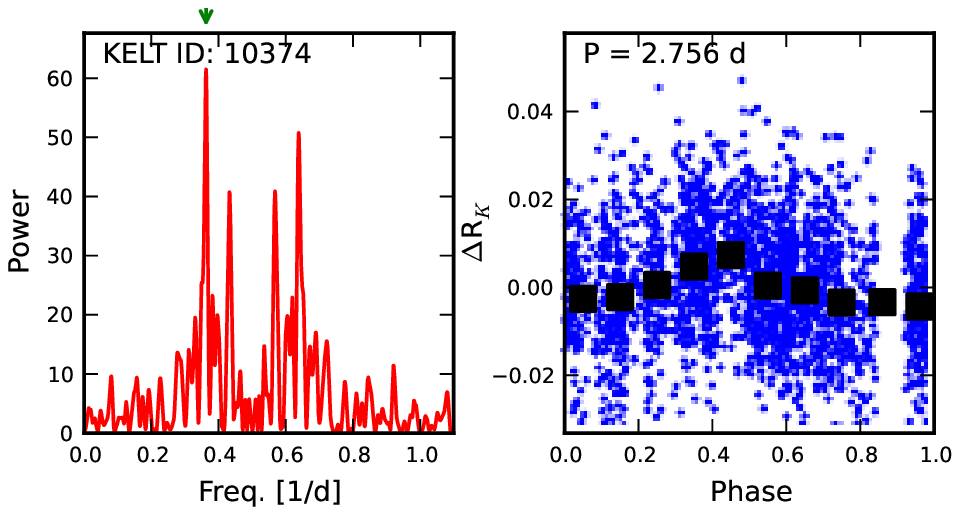}
	\includegraphics[scale=0.5,angle=0]{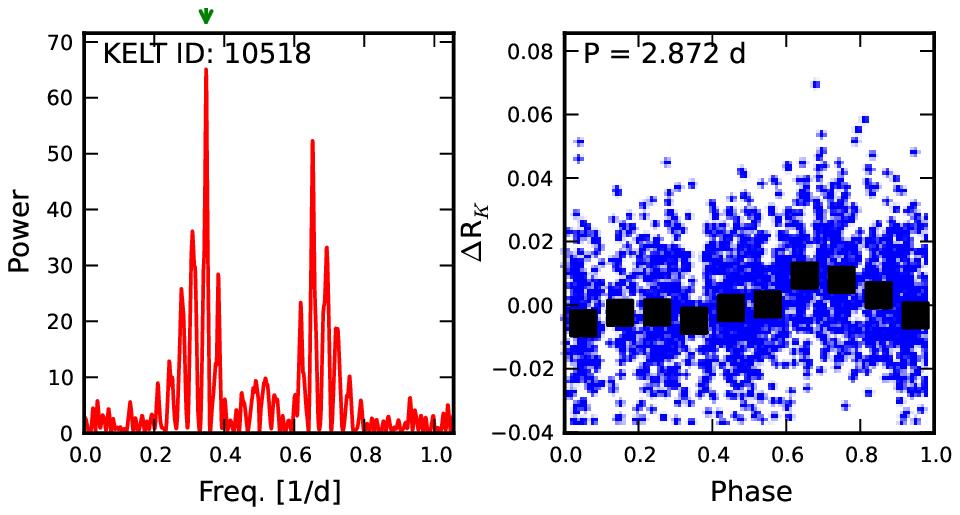}
	\includegraphics[scale=0.5,angle=0]{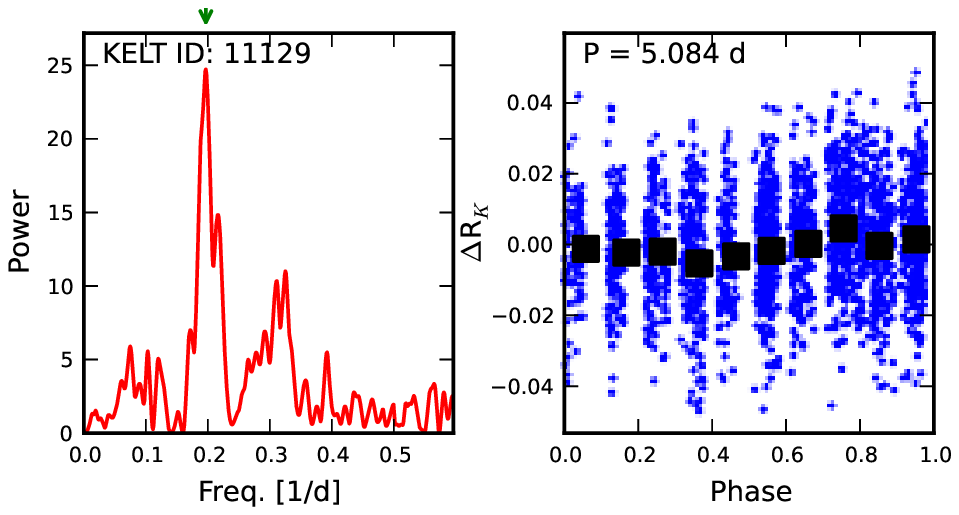}
	\includegraphics[scale=0.5,angle=0]{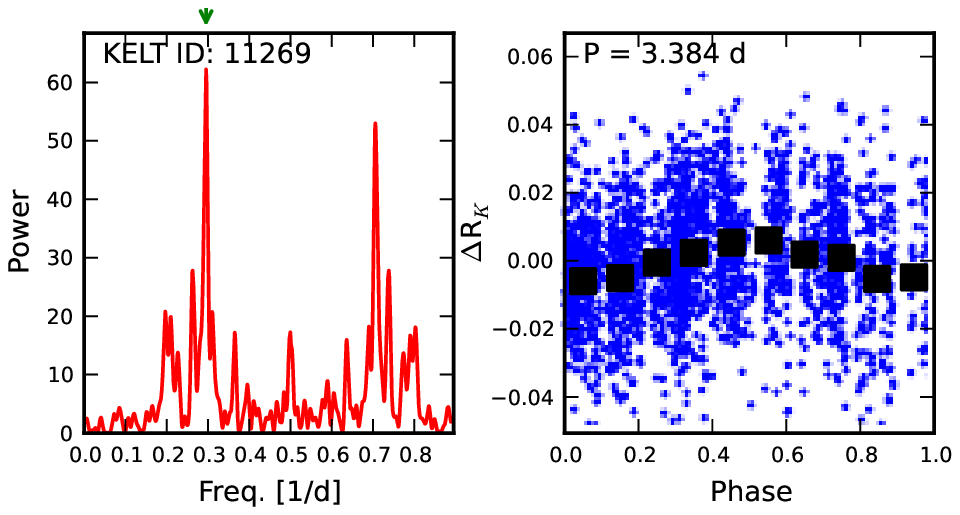}
	\includegraphics[scale=0.5,angle=0]{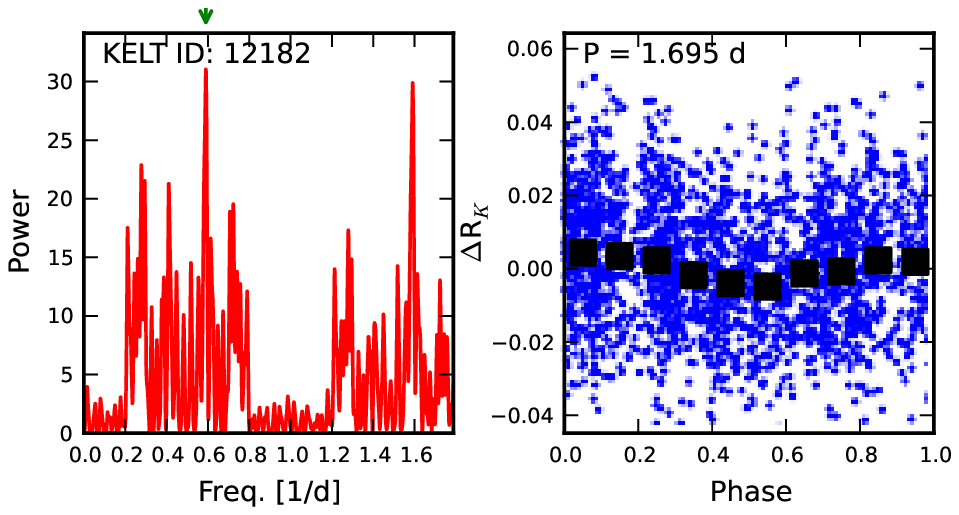}
	\includegraphics[scale=0.5,angle=0]{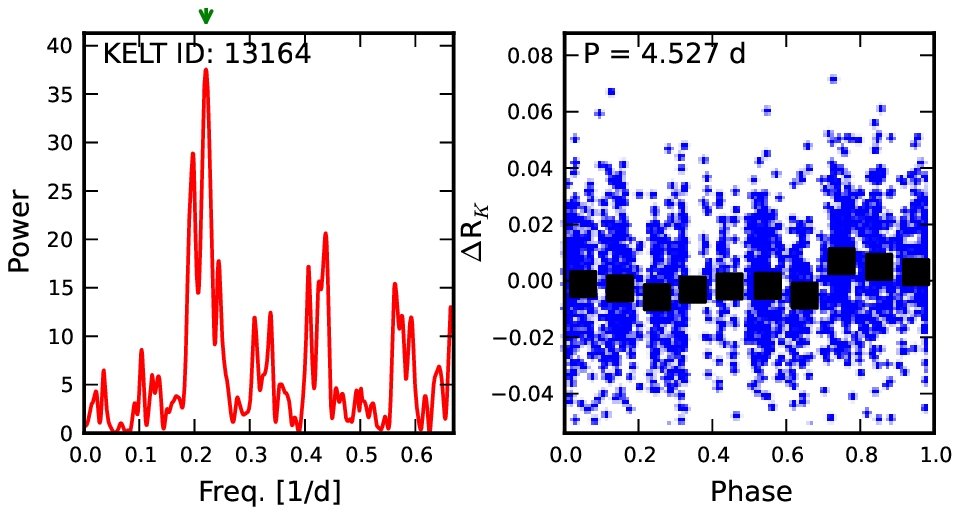}
	\includegraphics[scale=0.5,angle=0]{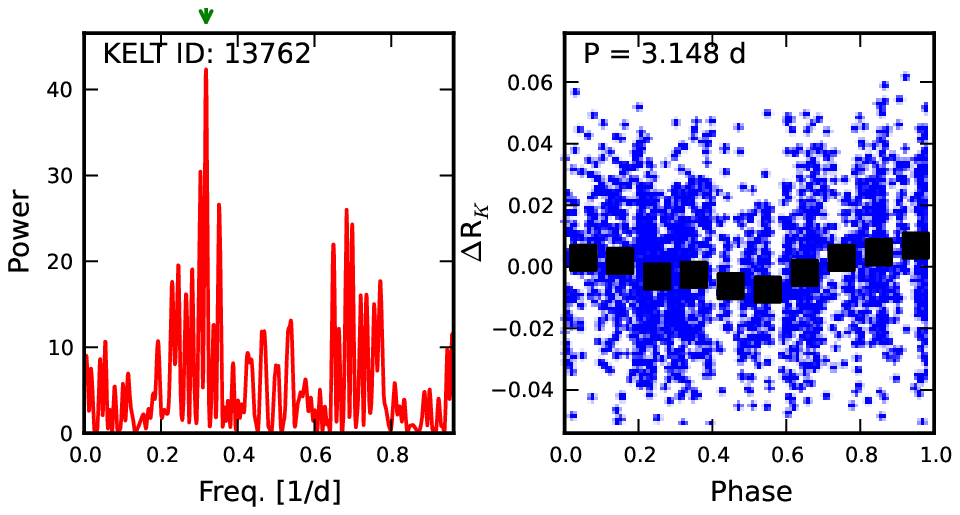}
	\includegraphics[scale=0.5,angle=0]{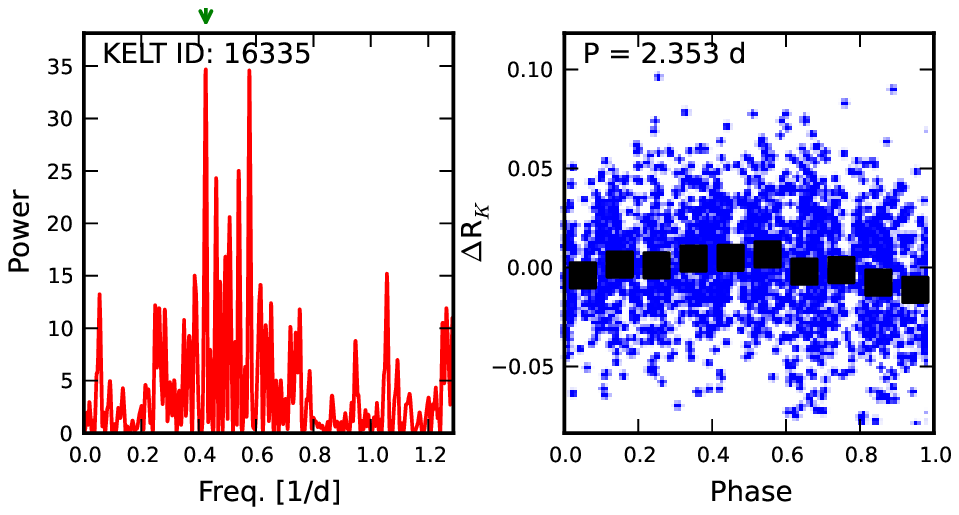}
	\includegraphics[scale=0.5,angle=0]{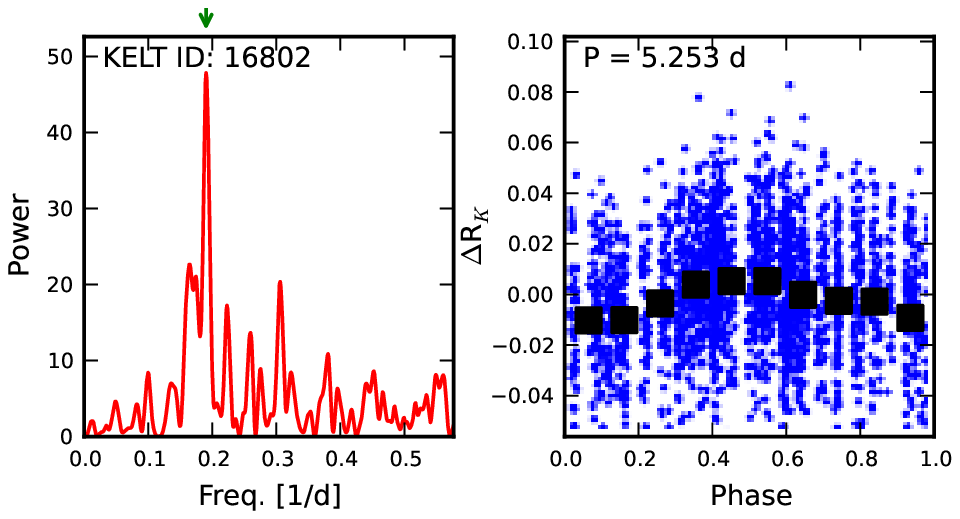}
	\caption{Lomb-Scargle periodograms and phased lightcurves for 35 Blanco~1 stars with measured periodicity in their KELT-South light curves.}
\end{figure}
\begin{figure}[!h]
 	\figurenum{14}
	\includegraphics[scale=0.5,angle=0]{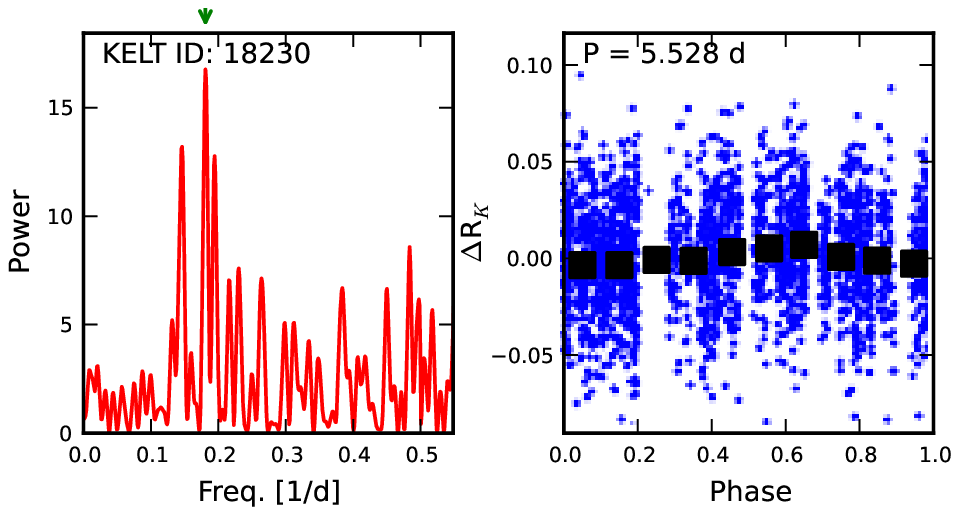}
	\includegraphics[scale=0.5,angle=0]{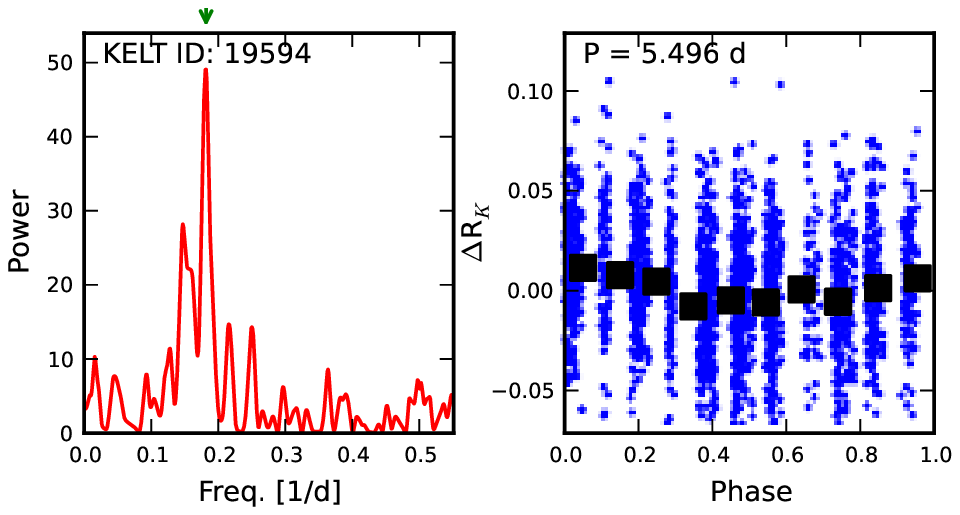}
	\includegraphics[scale=0.5,angle=0]{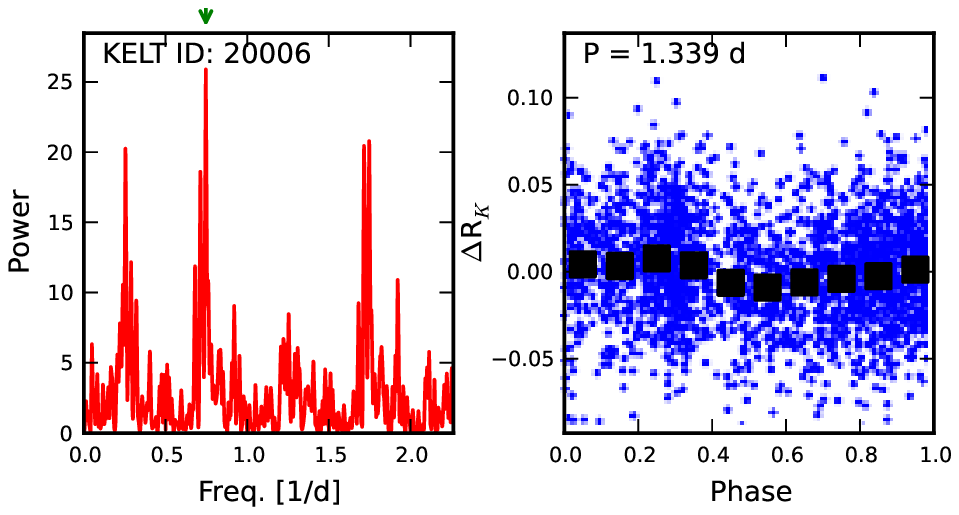}
	\includegraphics[scale=0.5,angle=0]{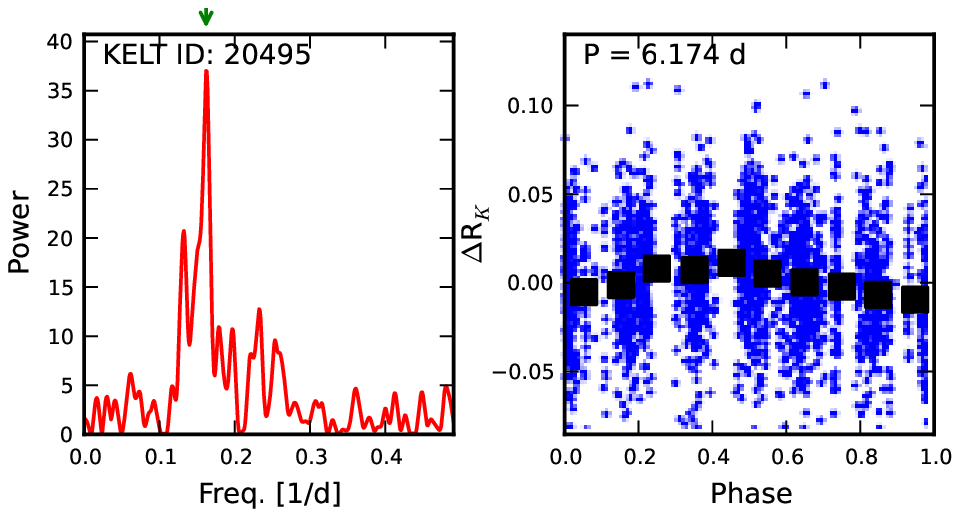}
	\includegraphics[scale=0.5,angle=0]{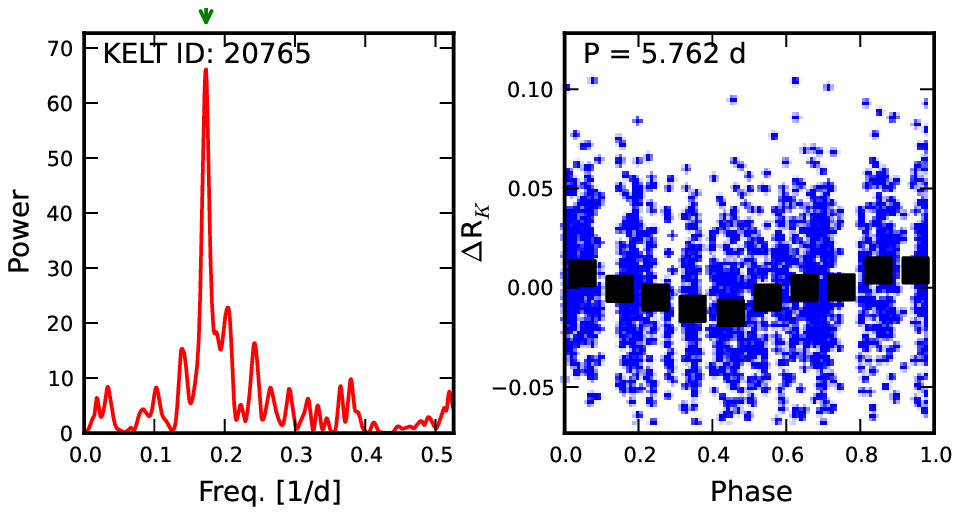}
	\includegraphics[scale=0.5,angle=0]{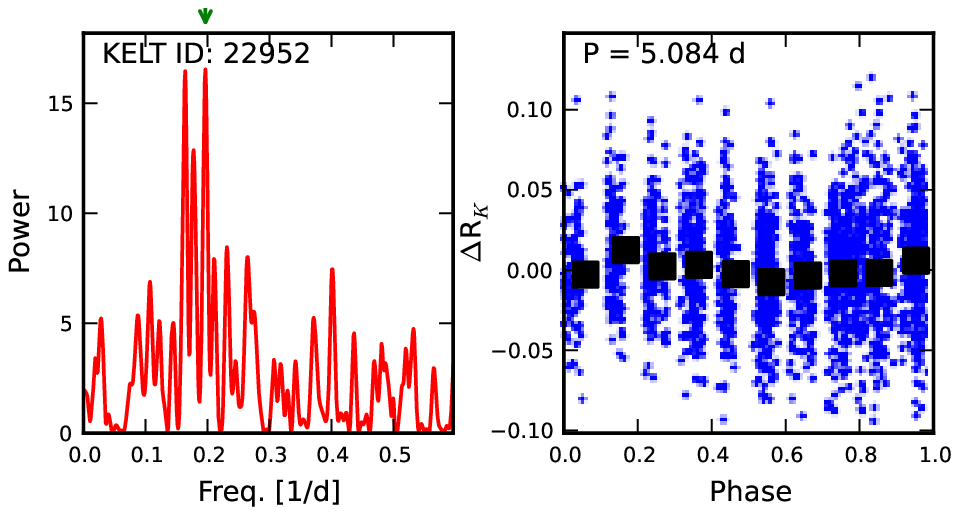}
	\includegraphics[scale=0.5,angle=0]{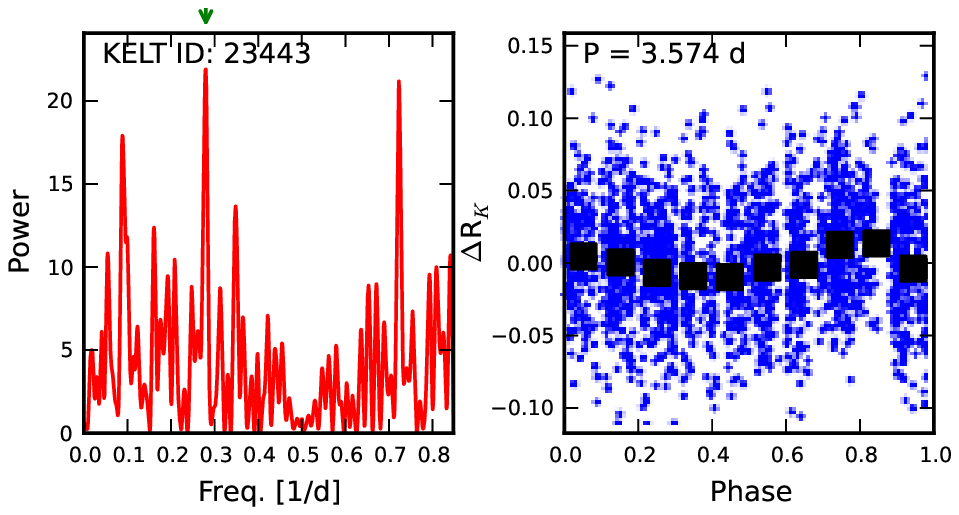}
	\includegraphics[scale=0.5,angle=0]{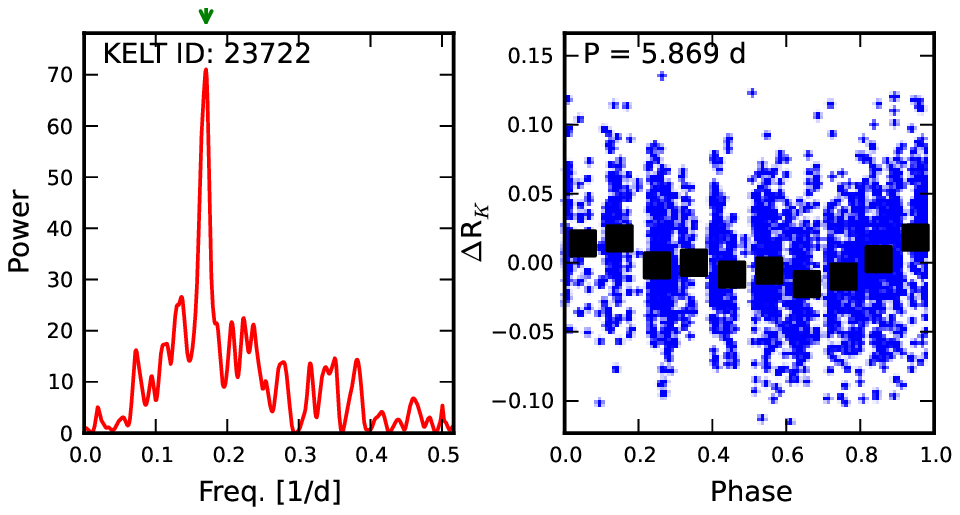}
	\includegraphics[scale=0.5,angle=0]{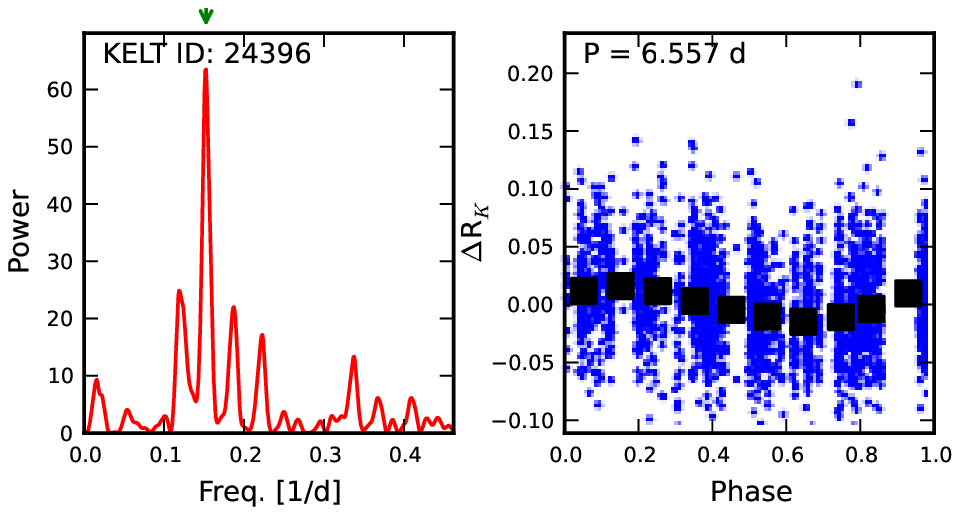}
	\includegraphics[scale=0.5,angle=0]{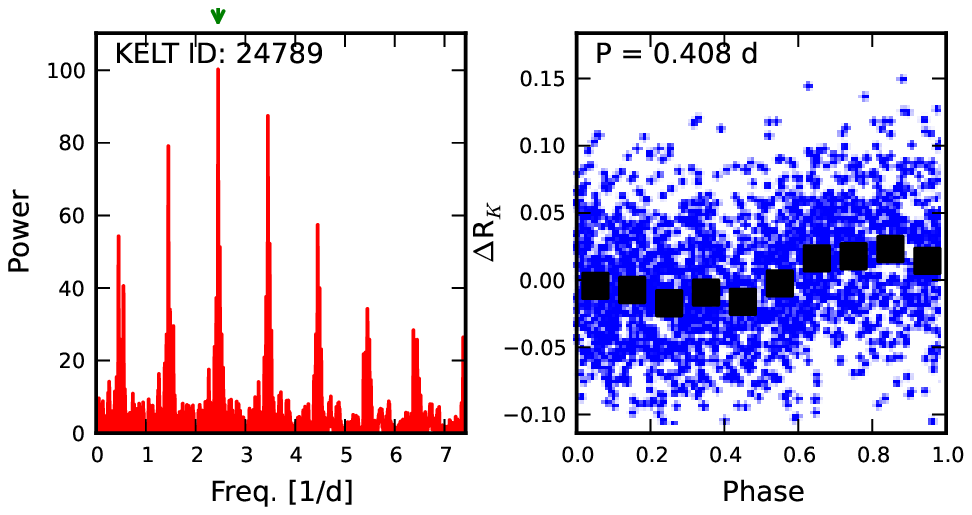}
	\includegraphics[scale=0.5,angle=0]{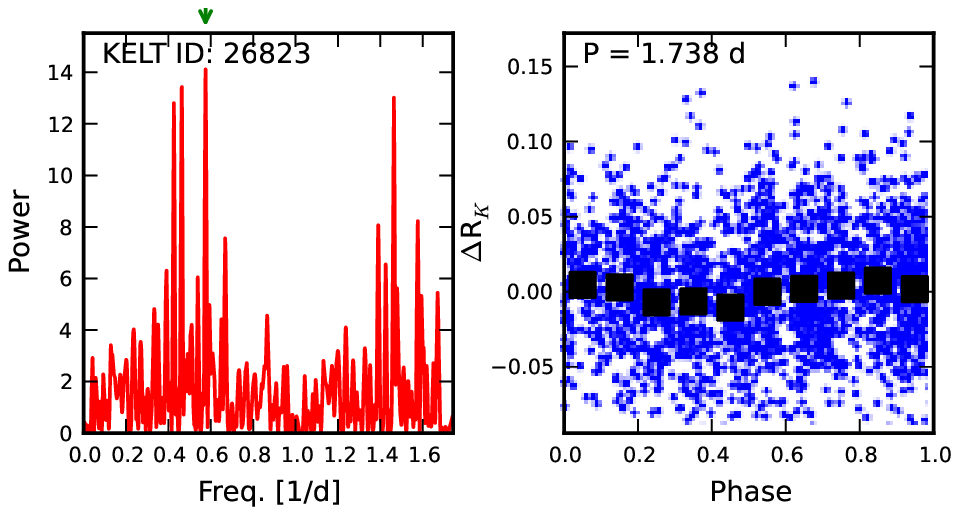}
	\includegraphics[scale=0.5,angle=0]{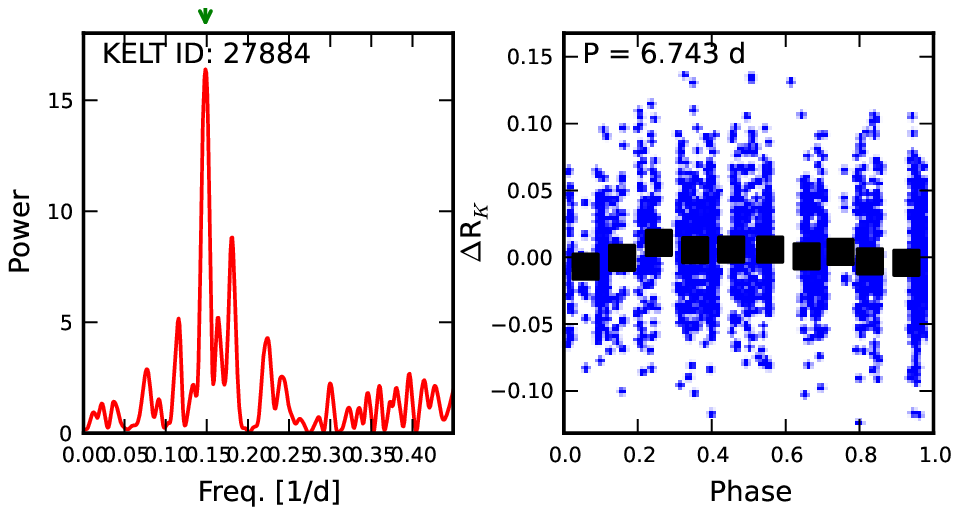}
	\includegraphics[scale=0.5,angle=0]{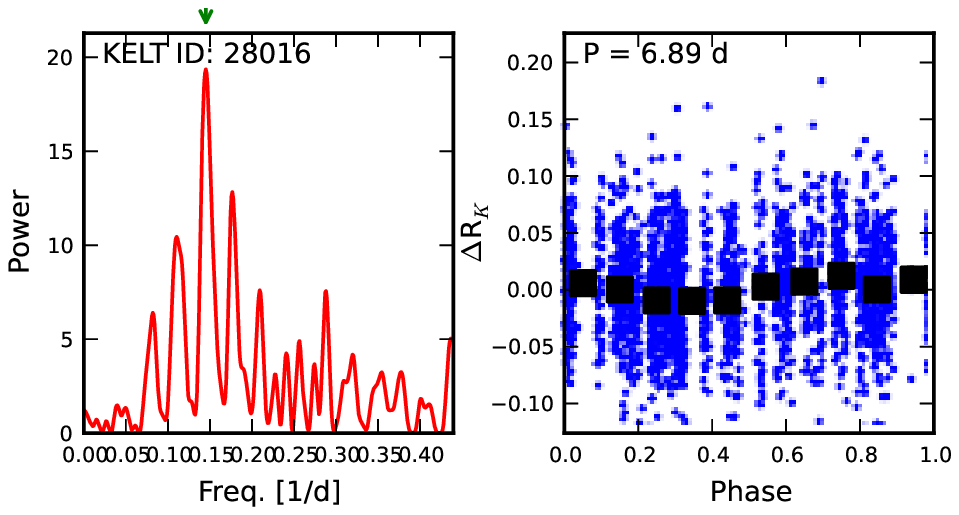}
	\includegraphics[scale=0.5,angle=0]{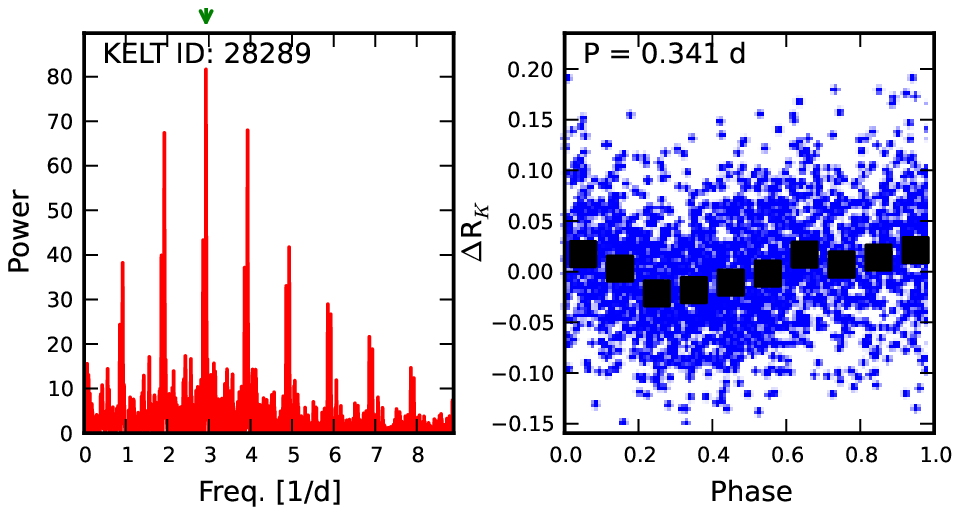}
	\includegraphics[scale=0.5,angle=0]{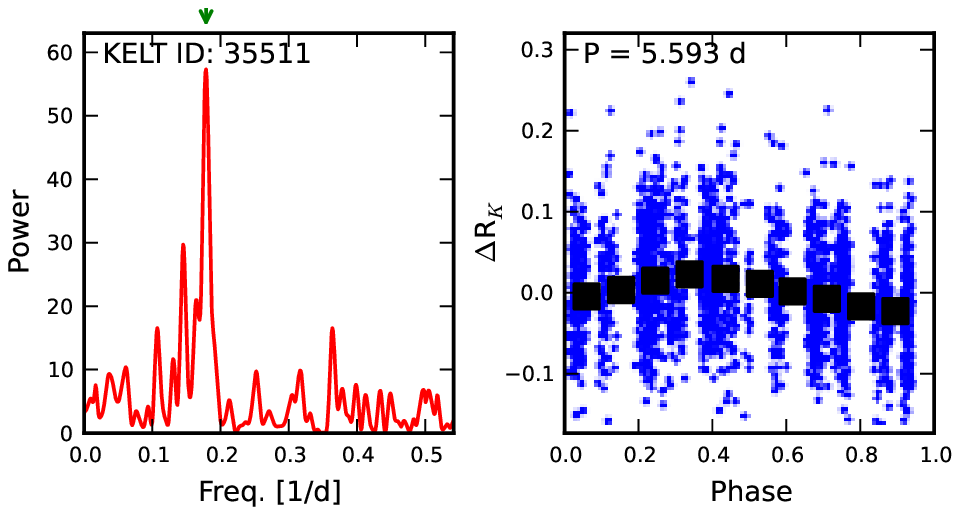}
	\includegraphics[scale=0.5,angle=0]{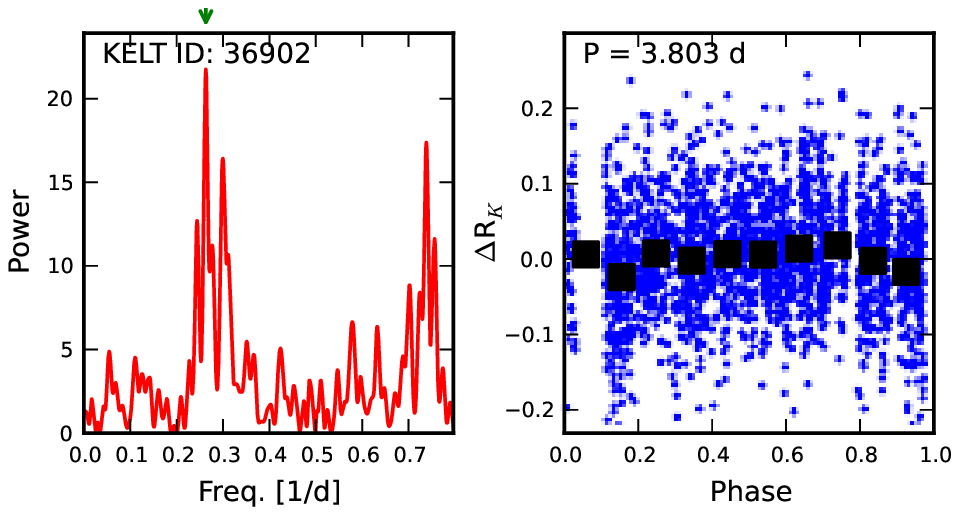}
	\includegraphics[scale=0.5,angle=0]{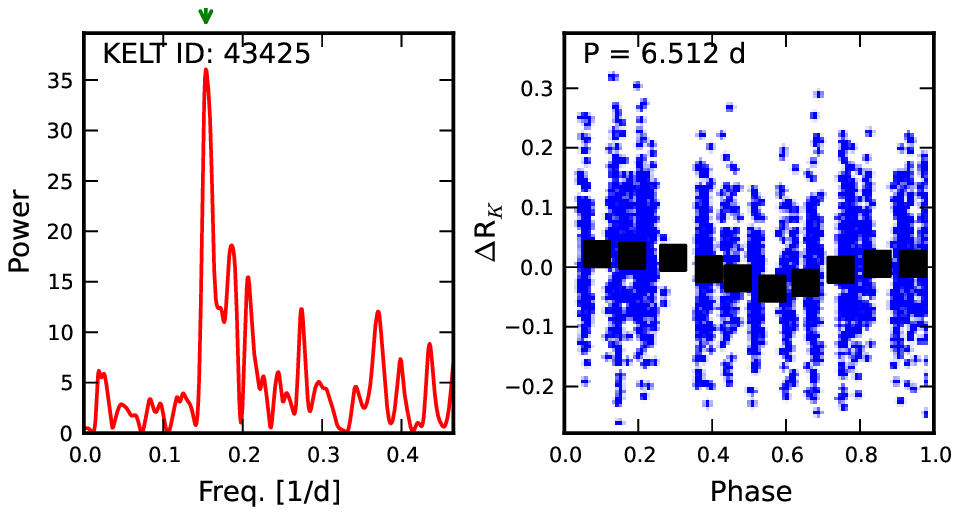}
	\caption{cont.}
\end{figure}

\clearpage
\bibliography{ms}

\begin{thebibliography}{57}
\expandafter\ifx\csname natexlab\endcsname\relax\def\natexlab#1{#1}\fi

\bibitem[{Barnes(2003)}]{Barnes2003a}
Barnes, S. 2003, \apj, 586, 464

\bibitem[{Barnes(2007)}]{Barnes2007}
---. 2007, \apj, 669, 1167

\bibitem[{Barnes(2010)}]{Barnes2010b}
---. 2010, \apj, 722, 222

\bibitem[{Barnes \& Kim(2010)}]{Barnes2010a}
Barnes, S., \& Kim, Y.-C. 2010, \apj, 721, 675

\bibitem[{Burke {et~al.}(2004)Burke, Pinsonneault, \& Sills}]{Burke2004}
Burke, C., Pinsonneault, M., \& Sills, A. 2004, \apj, 604, 272

\bibitem[{Cargile \& James(2010)}]{Cargile2010}
Cargile, P., \& James, D. 2010, \aj, 140, 677

\bibitem[{Cargile {et~al.}(2010)Cargile, James, \& Jeffries}]{Cargile2010b}
Cargile, P., James, D., \& Jeffries, R. 2010, \apj, 725, L111

\bibitem[{Cargile {et~al.}(2009)Cargile, James, \& Platais}]{Cargile2009}
Cargile, P., James, D., \& Platais, I. 2009, \aj, 137, 3230

\bibitem[{Casagrande {et~al.}(2010)Casagrande, Ram\'{\i}rez, Mel\'{e}ndez,
  Bessell, \& Asplund}]{Casagrande2010}
Casagrande, L., Ram\'{\i}rez, I., Mel\'{e}ndez, J., Bessell, M., \& Asplund, M.
  2010, \aap, 512, A54

\bibitem[{Casewell {et~al.}(2012)Casewell, Baker, Jameson, Hodgkin, Dobbie, \&
  Moraux}]{Casewell2012}
Casewell, S.~L., Baker, D. E.~A., Jameson, R.~F., Hodgkin, S.~T., Dobbie,
  P.~D., \& Moraux, E. 2012, \mnras, 425, 3112

\bibitem[{Chaboyer {et~al.}(1995)Chaboyer, Demarque, \&
  Pinsonneault}]{Chaboyer1995}
Chaboyer, B., Demarque, P., \& Pinsonneault, M.~H. 1995, \apj, 441, 865

\bibitem[{Charbonnel 
\& Talon(2005)}]{Charbonnel2005} Charbonnel, C., \& Talon, S.\ 2005, Science, 309, 2189 

\bibitem[{de~Epstein \& Epstein(1985)}]{deEpstein1985}
de~Epstein, A., \& Epstein, I. 1985, \aj, 90, 1211

\bibitem[{Denissenkov et al.(2010)}]{Denissenkov2010} Denissenkov, P.~A., 
Pinsonneault, M., Terndrup, D.~M., \& Newsham, G.\ 2010, \apj, 716, 1269 

\bibitem[{Epstein \& Pinsonneault(2012)}]{Epstein2012}
Epstein, C.~R., \& Pinsonneault, M.~H. 2012, arXiv:1203.1618

\bibitem[{Evans(1971)}]{Evans1971}
Evans, D.~S. 1971, \mnras, 154, 329

\bibitem[{Findeisen {et~al.}(2011)Findeisen, Hillenbrand, \&
  Soderblom}]{Findeisen2011}
Findeisen, K., Hillenbrand, L., \& Soderblom, D. 2011, \aj, 142, 23

\bibitem[{Ford {et~al.}(2005)Ford, Jeffries, \& Smalley}]{Ford2005}
Ford, A., Jeffries, R., \& Smalley, B. 2005, \mnras, 364, 272

\bibitem[{Foreman-Mackey {et~al.}(2013)Foreman-Mackey, Hogg, Lang, \&
  Goodman}]{Foreman-Mackey2013}
Foreman-Mackey, D., Hogg, D.~W., Lang, D., \& Goodman, J. 2013, \pasp, 125, 306

\bibitem[{Gallet 
\& Bouvier(2013)}]{Gallet2013} Gallet, F., \& Bouvier, J.\ 2013, \aap, 556, A36 

\bibitem[{Gonz\'{a}lez \& Levato(2009)}]{Gonzalez2009}
Gonz\'{a}lez, J., \& Levato, H. 2009, \aap, 507, 541

\bibitem[{Gregory {et~al.}(2012)Gregory, Donati, Morin, Hussain, Mayne,
  Hillenbrand, \& Jardine}]{Gregory2012}
Gregory, S., Donati, J., Morin, J., Hussain, G.~J., Mayne, N.~J., Hillenbrand,
  L.~A., \& Jardine, M. 2012, \apj, 755, 97

\bibitem[{Hartman {et~al.}(2010)Hartman, Bakos, Kov\'{a}cs, \&
  Noyes}]{Hartman2010}
Hartman, J., Bakos, G., Kov\'{a}cs, G., \& Noyes, R. 2010, \mnras, 408, 475

\bibitem[{Hartman {et~al.}(2009)Hartman, Gaudi, Pinsonneault, Stanek, Holman,
  McLeod, Meibom, Barranco, \& Kalirai}]{Hartman2009}
Hartman, J., {et~al.} 2009, \apj, 691, 342

\bibitem[{Henry {et~al.}(2011)Henry, Fekel, \& Henry}]{Henry2011}
Henry, G.~W., Fekel, F.~C., \& Henry, S.~M. 2011, \aj, 142, 39

\bibitem[{Herbst \& Mundt(2005)}]{Herbst2005}
Herbst, W., \& Mundt, R. 2005, \apj, 633, 967

\bibitem[{Irwin 
\& Bouvier(2009)}]{Irwin2009} Irwin, J., \& Bouvier, J.\ 2009, IAU Symposium, 258, 363 

\bibitem[{James {et~al.}(2010)James, Barnes, Meibom, Lockwood, Levine,
  Deliyannis, Platais, Steinhauer, \& Hurley}]{James2010}
James, D., {et~al.} 2010, \aap, 515, A100

\bibitem[{Jeffries \& James(1999)}]{Jeffries1999a}
Jeffries, R., \& James, D. 1999, \apj, 511, 218

\bibitem[{Kawaler(1988)}]{Kawaler1988}
Kawaler, S. 1988, \apj, 333, 236

\bibitem[{Kim \& Demarque(1996)}]{Kim1996}
Kim, Y.-C., \& Demarque, P. 1996, \apj, 457, 340

\bibitem[{Kov\'{a}cs {et~al.}(2005)Kov\'{a}cs, Bakos, \& Noyes}]{Kovacs2005}
Kov\'{a}cs, G., Bakos, G., \& Noyes, R.~W. 2005, \mnras, 356, 557

\bibitem[{Kron(1947)}]{Kron1947}
Kron, G.~E. 1947, \pasp, 59, 261

\bibitem[{Krzeminski(1969)}]{Krzeminski1969}
Krzeminski, W. 1969, Low-Luminosity Stars

\bibitem[{Mamajek \& Hillenbrand(2008)}]{Mamajek2008}
Mamajek, E., \& Hillenbrand, L. 2008, \apj, 687, 1264

\bibitem[{Martin {et~al.}(2005)Martin, Fanson, Schiminovich, Morrissey,
  Friedman, Barlow, Conrow, Grange, Jelinsky, Milliard, Siegmund, Bianchi,
  Byun, Donas, Forster, Heckman, Lee, Madore, Malina, Neff, Rich, Small,
  Surber, Szalay, Welsh, \& Wyder}]{Martin2005}
Martin, D.~C., {et~al.} 2005, \apj, 619, L1

\bibitem[{Matt {et~al.}(2012)Matt, MacGregor, Pinsonneault, \&
  Greene}]{Matt2012}
Matt, S.~P., MacGregor, K.~B., Pinsonneault, M.~H., \& Greene, T.~P. 2012,
  \apj, 754, L26

\bibitem[{Meibom {et~al.}(2009)Meibom, Mathieu, \& Stassun}]{Meibom2009}
Meibom, S., Mathieu, R., \& Stassun, K. 2009, \apj, 695, 679

\bibitem[{Meibom {et~al.}(2011{\natexlab{a}})Meibom, Mathieu, Stassun,
  Liebesny, \& Saar}]{Meibom2011}
Meibom, S., Mathieu, R., Stassun, K., Liebesny, P., \& Saar, S.
  2011{\natexlab{a}}, \apj, 733, 115

\bibitem[{Meibom {et~al.}(2011{\natexlab{b}})Meibom, Barnes, Latham, Batalha,
  Borucki, Koch, Basri, Walkowicz, Janes, Jenkins, {Van Cleve}, Haas, Bryson,
  Dupree, Furesz, Szentgyorgyi, Buchhave, Clarke, Twicken, \&
  Quintana}]{Meibom2011b}
Meibom, S., {et~al.} 2011{\natexlab{b}}, \apj, 733, L9

\bibitem[{Mermilliod {et~al.}(2008)Mermilliod, Platais, James, Grenon, \&
  Cargile}]{Mermilliod2008a}
Mermilliod, J.-C., Platais, I., James, D., Grenon, M., \& Cargile, P. 2008,
  \aap, 485, 95

\bibitem[{Micela {et~al.}(1999)Micela, Sciortino, Favata, Pallavicini, \&
  Pye}]{Micela1999a}
Micela, G., Sciortino, S., Favata, F., Pallavicini, R., \& Pye, J. 1999, \aap,
  344, 83

\bibitem[{Moraux {et~al.}(2007)Moraux, Bouvier, Stauffer, {Barrado y
  Navascu\'{e}s}, \& Cuillandre}]{Moraux2007}
Moraux, E., Bouvier, J., Stauffer, J., {Barrado y Navascu\'{e}s}, D., \&
  Cuillandre, J.-C. 2007, \aap, 471, 499

\bibitem[{Pepper {et~al.}(2012)Pepper, Kuhn, Siverd, James, \&
  Stassun}]{Pepper2012}
Pepper, J., Kuhn, R., Siverd, R., James, D., \& Stassun, K. 2012, \pasp, 124,
  230

\bibitem[{Pickering(1881)}]{Pickering1881}
Pickering, E.~C. 1881, Proc. Amer. Acad. Arts and Sci., 16

\bibitem[{Pillitteri {et~al.}(2003)Pillitteri, Micela, Sciortino, \&
  Favata}]{Pillitteri2003}
Pillitteri, I., Micela, G., Sciortino, S., \& Favata, F. 2003, \aap, 399, 919

\bibitem[{Platais {et~al.}(2011)Platais, Girard, Vieira, L\'{o}pez, Loomis,
  McLean, Pourbaix, Moraux, Mermilliod, James, Cargile, Barnes, \&
  Castillo}]{Platais2011}
Platais, I., {et~al.} 2011, \mnras, 413, 1024

\bibitem[{Press \& Rybicki(1989)}]{Press1989}
Press, W.~H., \& Rybicki, G.~B. 1989, \apj, 338, 277

\bibitem[{Reiners \& Mohanty(2012)}]{Reiners2012}
Reiners, A., \& Mohanty, S. 2012, \apj, 746, 43

\bibitem[{Schwarzenberg-Czerny(1991)}]{Schwarzenberg-Czerny1991}
Schwarzenberg-Czerny, A. 1991, \mnras, 253, 198

\bibitem[{Scott(1992)}]{Scott1992}
Scott, D.~W. 1992, Multivariate Density Estimation

\bibitem[{Silverman(1986)}]{Silverman1986}
Silverman, B.~W. 1986, Monographs on Statistics and Applied Probability

\bibitem[{Siverd {et~al.}(2012)Siverd, Beatty, Pepper, Eastman, Collins,
  Bieryla, Latham, Buchhave, Jensen, Crepp, Street, Stassun, {Scott Gaudi},
  Berlind, Calkins, DePoy, Esquerdo, Fulton, Fűr\'{e}sz, Geary, Gould, Hebb,
  Kielkopf, Marshall, Pogge, Stanek, Stefanik, Szentgyorgyi, Trueblood,
  Trueblood, Stutz, \& van Saders}]{Siverd2012}
Siverd, R.~J., {et~al.} 2012, \apj, 761, 123

\bibitem[{Skumanich(1972)}]{Skumanich1972}
Skumanich, A. 1972, \apj, 171, 565

\bibitem[{Soderblom {et~al.}(2005)Soderblom, Nelan, Benedict, McArthur,
  Ramirez, Spiesman, \& Jones}]{Soderblom2005}
Soderblom, D., Nelan, E., Benedict, G., McArthur, B., Ramirez, I., Spiesman,
  W., \& Jones, B. 2005, \aj, 129, 1616

\bibitem[{Soderblom {et~al.}(2009)Soderblom, Laskar, Valenti, Stauffer, \&
  Rebull}]{Soderblom2009}
Soderblom, D.~R., Laskar, T., Valenti, J.~A., Stauffer, J.~R., \& Rebull, L.~M.
  2009, \aj, 138, 1292

\bibitem[{Soderblom(2010)}]{Soderblom2010}
Soderblom, D. R.~D. 2010, \araa, 48, 581

\bibitem[{Spada et al.(2011)}]{Spada2011} Spada, F., Lanzafame, 
A.~C., Lanza, A.~F., Messina, S., 
\& Collier Cameron, A.\ 2011, \mnras, 416, 447 

\bibitem[{Stauffer {et~al.}(2007)Stauffer, Hartmann, Fazio, Allen, Patten,
  Lowrance, Hurt, Rebull, Cutri, Ramirez, Young, Rieke, Gorlova, Muzerolle,
  Slesnick, \& Skrutskie}]{Stauffer2007}
Stauffer, J., {et~al.} 2007, \apjs, 172, 663

\bibitem[{Tognelli {et~al.}(2011)Tognelli, {Prada Moroni}, \&
  Degl’Innocenti}]{Tognelli2011}
Tognelli, E., {Prada Moroni}, P.~G., \& Degl’Innocenti, S. 2011, \aap, 533,
  A109

\bibitem[{van Leeuwen(2009)}]{vanLeeuwen2009}
van Leeuwen, F. 2009, \aap, 497, 209

\bibitem[{Westerlund {et~al.}(1988)Westerlund, Lundgren, Pettersson, Garnier,
  \& Breysacher}]{Westerlund1988}
Westerlund, B., Lundgren, K., Pettersson, B., Garnier, R., \& Breysacher, J.
  1988, \aaps, 76, 101

\end{thebibliography}

\end{document}